\documentclass[aps,pre,subeqns.floatfix,amsmath,onecolumn]{revtex4}
\usepackage[pdftex]{graphicx}
\usepackage{amsmath}
\usepackage{amsfonts}
\usepackage{graphicx}
\usepackage{hyperref}
\usepackage{comment}
\usepackage{mathtools}
\usepackage{verbatim}
\usepackage[outdir=./]{epstopdf}
\usepackage{subfigure}
\usepackage{hyperref}
\usepackage{latexsym}
\usepackage{dcolumn}
\usepackage{epsf}
\usepackage{float}

\usepackage{color}
\usepackage[utf8]{inputenc}
\begin{document}
\title{A double explosive Kuramoto transition in hypergraphs}
\author{Sangita Dutta$^1$}
\email{sangitaduttaprl@gmail.com}
\author{Prosenjit Kundu$^2$}
\author{Pitambar Khanra$^3$}
\author{Ludovico Minati$^{4,5}$}
\email{lminati@uestc.edu.cn; lminati@ieee.org}
\author{Stefano Boccaletti$^{6,7,8}$}
\author{Pinaki Pal$^1$}
\email{ppal.maths@nitdgp.ac.in}
\author{Chittaranjan Hens$^9$}
\affiliation{$^1$Department of Mathematics, National Institute of Technology, Durgapur~713209, India}
\affiliation{$^2$Dhirubhai Ambani Institute of Information and Communication Technology, Gandhinagar, Gujarat, 382007, India}
\affiliation{$^3$Department of Chemical and Biological Engineering, The State University of New York at Buffalo, Buffalo, New York 14260, USA}
\affiliation{$^4$School of Life Science and Technology, University of Electronic Science and Technology of China, 611731 Chengdu, China}
\affiliation{$^5$Center for Mind/Brain Sciences (CIMeC), University of Trento, 38123 Trento, Italy}
\affiliation{$^6$Research Institute of Interdisciplinary Intelligent Science, Ningbo University of Technology, 315104 China}
\affiliation{$^7$Sino-Europe Complexity Science Center, North University of China, 030051, Taiyuan, China}
\affiliation{$^8$CNR - Institute of Complex Systems, Via Madonna del Piano 10, I-50019, Sesto Fiorentino, Italy}
\affiliation{$^9$Center for Computational Natural Science and Bioinformatics, International Institute of Informational Technology, Gachibowli, Hyderabad 500032, India}

\begin{abstract}
This study aims to develop a generalised concept that will enable double explosive transitions in the forward and backward directions or a combination thereof. We found two essential factors for generating such phase transitions: the use of higher-order (triadic) interactions and the partial adaptation of a global order parameter acting on the triadic coupling.
A compromise between the two factors may result in a double explosive transition. To reinforce numerical observations, we employed the Ott--Antonsen ansatz. We observed that for a wide class of hypergraphs, combining two elements can result in a double explosive transition. 
\end{abstract}
\maketitle
\section{Introduction}
\par Classical network theory states that if a link joins two nodes with a probability $p$, a continuous percolation process will develop a giant connected component (GCC). An analytically tractable critical $p$ \cite{Newman_Network_Book, Erdos_PubMathDeb1959} constitutes the foundation for such GCC emergence. Later research showed that if a specific product rule is applied at the time of connection, the critical $p$ can be delayed and the transition becomes explosive (Achlioptas processes; two links compete to be added) \cite{li2021percolation,achlioptas2009explosive}. More complex behaviour can also be guaranteed by further alterations to the connection rule; specifically, by using the three-vertex rule, which connects nodes based on their cluster sizes. This enables numerous discontinuous jumps in the arbitrary vicinity of the first continuous jump \cite{nagler2012continuous, d2015explosive}.
\par Non-equilibrium phase transitions are frequently seen in the networks of complex systems such as Kuramoto oscillators. In particular, the synchronisation transitions that characterise the path from incoherence to coherence are determined by the coupling configuration and distribution of intrinsic natural frequencies \cite{boccaletti2014structure, rodrigues2016kuramoto, kuramoto1984chemical, kundu2018perfect, zanin2016combining}. For example, if the natural frequencies and network structure are correlated and the network is highly heterogeneous \cite{coutinho2013kuramoto}, the transition becomes explosive \cite{gomez2011explosive, qiu2015landau}. Another interesting technique for explosive or discontinuous transitions is adapting the coupling with the Kuramoto order parameter \cite{Zhang_PRL2015, manoranjani2023phase, filatrella2007generalized, khanra2020amplification, khanra_chaossoli2021,biswas2024effect}. Such adaptations are made elegant by the following characteristics:(i) A network specification is not required. (ii) Different types of frequency sets can be operated across.
\par These two observations—one from networks (stair-like discontinuities) and the other from coupled Kuramoto oscillators (explosive transitions)—motivate the crucial question of whether the strategic creation of a method for producing stair-like (specifically, double-explosive-transition single steps) discontinuities in coupled Kuramoto oscillators is possible. If so, what kind of coupling will it have? Can we move it in only one direction, that is, forward or backward? Stair-like behaviour has been observed in the Sakaguchi--Kuramoto model in random networks with partial degree--frequency correlation \cite{kundu2019synchronization}.
Notably, the stair-like behaviour was observed in a finite graph of second-order Kuramoto oscillators. Varying sizes of clusters coexist in hysteric regimes when there is considerable inertia \cite{olmi2014hysteretic}. Further, more intricate stair-like behaviour was developed in \cite{gao2021synchronized} in the presence of high inertia and in \cite{carballosa2023cluster} in the presence of higher-order interactions. A recent study suggested that if suitable frequency selections are made and the network connection is temporally adjusted—such as with the Hebbian rule—a finite network of Kuramoto oscillators may display a stair-like discontinuous transition \cite{fialkowski2023heterogeneous}. Recently, a double hysteresis loop in duplex networks with a high phase lag has been quantitatively detected \cite{seif2024double}. However, to date, no universal mechanism or technique has been developed to realise a double hysteresis loop or explosive transition (with Kuramoto oscillators) in a single direction.
\par In this paper, we propose a broad approach that can produce a double explosive transition in one or both directions. Our method utilises two key factors: (i) We employed hypergraphs to account for pairwise and triadic interaction \cite{courtney2016generalized, landry2020effect}. (ii) Building on previous research \cite{Zhang_PRL2015}, we suggest a partial-coupling adaptation of the order parameter. The first factor guarantees tiered or explosive synchronisation (weak followed by explosive synchronisation in one direction and explosive synchronisation in the opposite direction) with or without adaptation \cite{skardal2020higher, skardal2022tiered, suman2024finite, dutta2023perfect, dutta2023impact, dutta2024transition}. The second factor enables control of the explosion width \cite{Zhang_PRL2015} without need to consider any particular networks or frequencies. We have observed multiple non-trivial incoherence to coherence transitions in a hypergraph of Kuramoto oscillators when both factors were combined. Based on the Ott--Antosen ansatz \cite{ott2008low}, for a hypergraph, we may establish that a partial adaptation can be used to increase or decrease the size of double explosive loops (for a given higher-order interaction strength). Evidence suggests that such double loops may be removed and a single stair can appear in either direction with careful adaptation. Ultimately, using average-frequency analysis, we demonstrate how the dyadic coupling strength—or the adaptation in triadic coupling—influences the development or decline of these clusters. %In the near future, we will focus on multiple stairs.
The coupled Kuramoto model can be expressed as %\section{Synchronization on Hypergraphs}
%\section{Analytic Derivation}
%The Kuramoto model regarding this hypergraph structure can be written as      
\begin{eqnarray}
\label{model_network}
\dot{\theta_i}=\omega_i+ K_1 r_1^a \sum_{j=1}^{N} A_{ij} \sin(\theta_j-\theta_i-\beta)+K_2r_1^b \sum_{j=1}^N \sum_{k=1}^N B_{ijk} \sin(2\theta_j-\theta_k-\theta_i-\beta), \hspace*{0.5cm} i=1,2,\dots,N. 
\end{eqnarray}
where $A_{ij}$
is the $ij$-th element of the adjacency matrix $A$
wherein $A_{ij}=1$ if there is a pairwise connection between the 
$i$-th and $j$-th node, otherwise$A_{ij}=0$; 
similarly, 
$B_{ijk}$ represents the triangular connections between the three nodes 
$i,~j$ and $k$,
wherein $B_{ijk}=1$ if the three nodes form a triangle, otherwise $0$;
$\omega_i$ is the natural frequency of the Lorentzian distribution $g(\omega)=\frac{\Delta}{\pi[\Delta^2+(\omega-\omega_0^2)]}$, wherein 
$\Delta$ is the half width and 
$\omega_0$ is the mean of the distribution;
$N$ is the total number of oscillators in the system and
the non-zero $a$ and $b$ control the strength of adaptation in pairwise (strength $K_1$) and higher-order coupling ($K_2$), respectively.
\begin{figure}[h!]
    \centering
    \includegraphics[width=9cm]{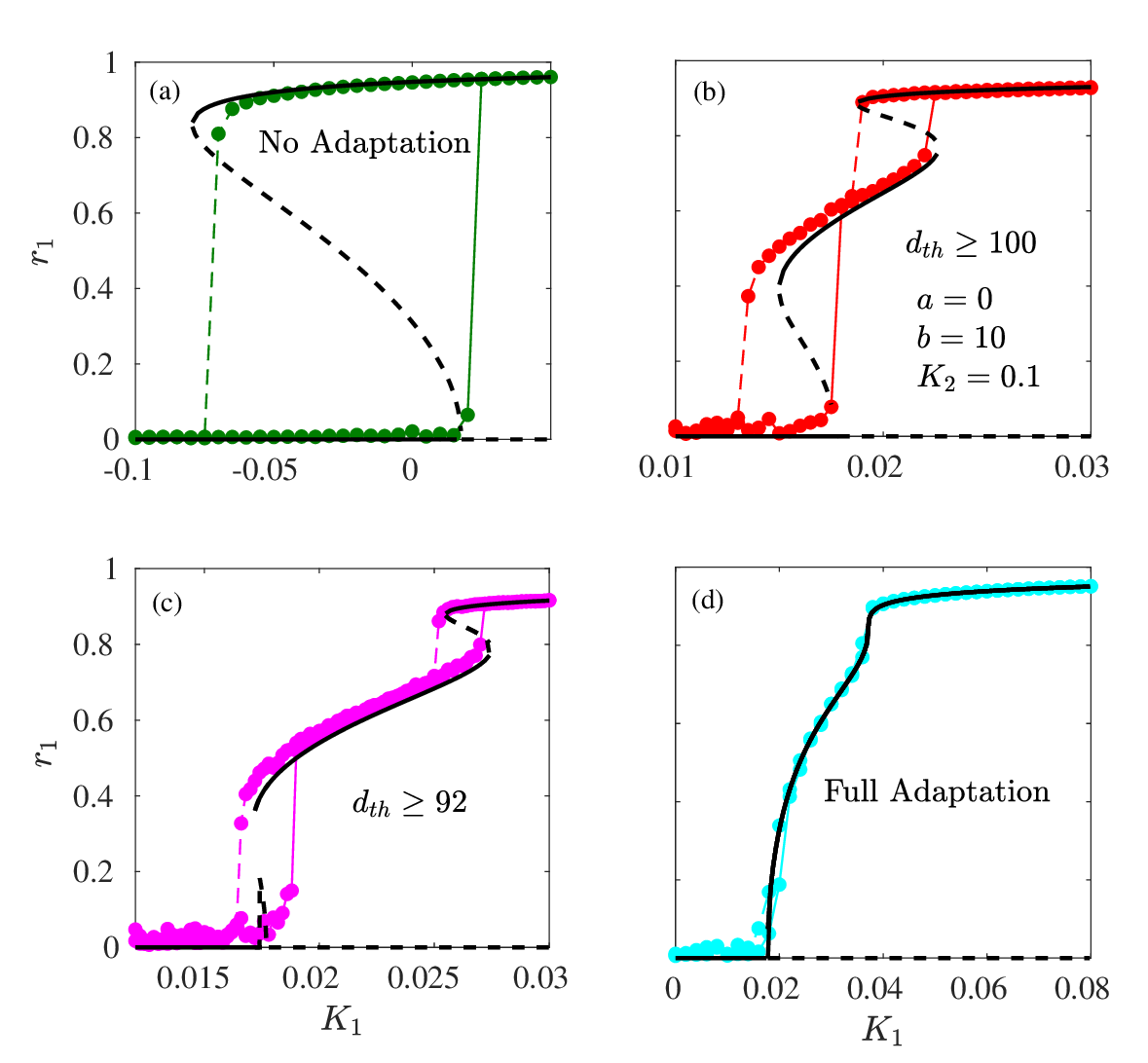}
    \caption{Synchronisation profiles showing $r_1$ as a function of $K_1$ for (a) no adaptation, (b) $d_{\rm th}=100$, (c) $d_{\rm th}=92$ and (d) full adaptation. Black solid and dashed lines indicate stable and unstable solutions of the self-consistent equations, respectively. Coloured (Green, red, magenta and cyan) full circles joined with solid and dashed lines represent numerically simulated data points for the forward and backward directions, respectively. All other parameters were fixed ($a=0,~b=10$ and $K_2=0.1$).
    }
    \label{UD_adaptK2_diff_percetage}
\end{figure}
$\beta$ indicates the phase frustration or phase lag of the system \cite{ichinomiya2004frequency, sakaguchi1986soluble, kundu2017transition}. We did not investigate the role of $\beta$ in this study; however it has been used in the analytical computation. Given that $\beta$ suppresses synchronisation \cite{kundu2017transition,dutta2024transition,dutta2023impact,omel2012nonuniversal}, %Chinese_scientific_reports 
conducting a separate study to determine how to create and destroy double or stair-like explosive synchronisation considering 
$\beta$ is necessary.  % {\color{red} Here the delta function ....}
For the construction of hypergraphs, please see the Supplemental Material, Sec.\ V.  
To proceed further, let us first define the order parameters, which characterise the synchronisation level. The local order parameters are defined as
%\begin{eqnarray}
%&&
$ R_i^1=\sum_{j=1}^{N}A_{ij}e^{i\theta_j}$,  %\nonumber 
and 
$ R_i^2=\sum_{j,k=1}^{N}B_{ijk}e^{2i\theta_j}e^{-i\theta_k}$
%\label{local_order_parameter}
%\end{eqnarray}
and the global order parameters can be defined as \cite{adhikari2023synchronization}
%\begin{eqnarray}
  $  z_1=r_1(t)e^{i\psi_1}=\frac{1}{N\langle k^{(1)}\rangle}\sum_{i=1}^N R_i^1$, and  %\nonumber\\
   $ z_2=r_2(t)e^{i\psi_2}=\frac{1}{2N\langle k^{(2)}\rangle}\sum_{i=1}^N R_i^2$, % \nonumber\\
    \label{global_order_parameter}
%\end{eqnarray}
where $\langle k^{(1)}\rangle$ and $\langle k^{(2)}\rangle$ denote the mean pairwise and triadic degrees, respectively, and $\psi_1$ and $\psi_2$ are the average phase velocities. In this study, the global order parameter $r_1$ was modified to consider both coupling strengths in the form of $K_1r_1^{a}$ and $K_2r_1^{b}$. 
\begin{figure*}%[h!]
    \centering
    \includegraphics[width=16cm]{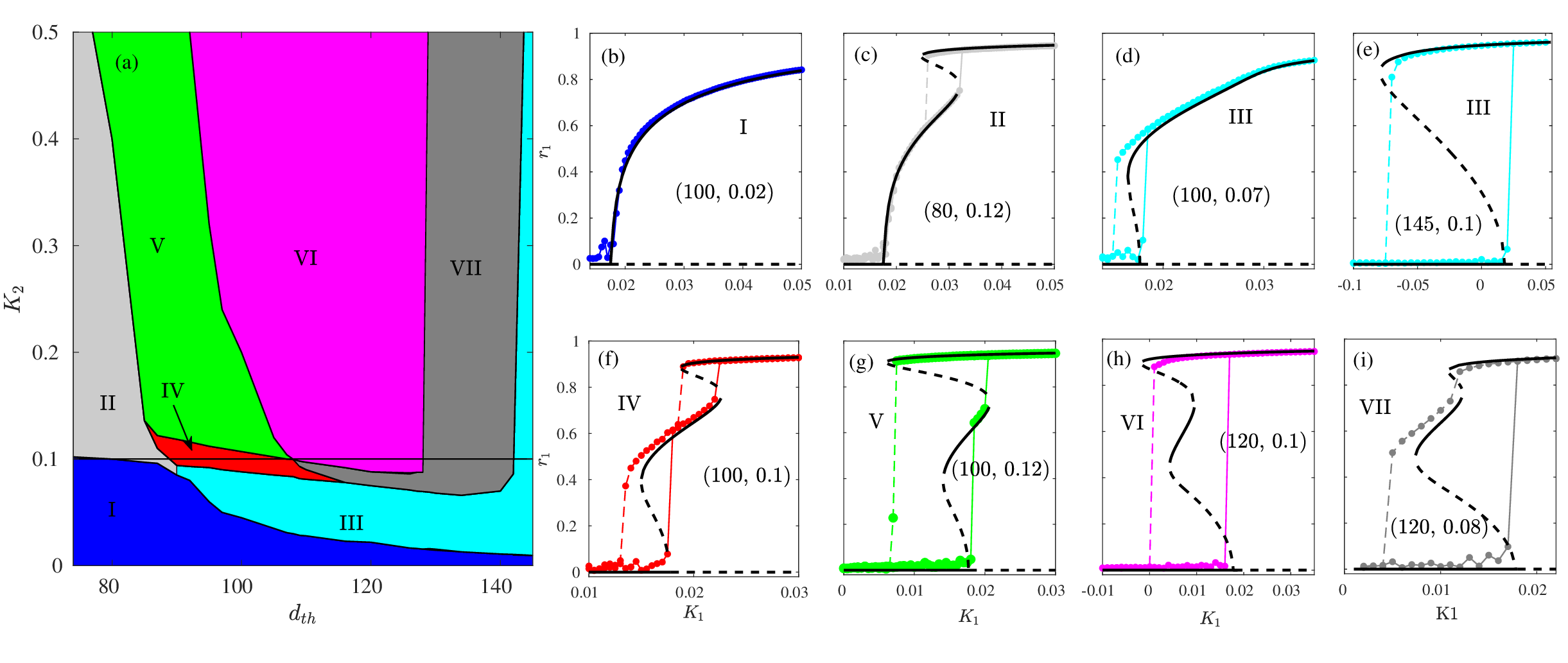}
    \caption{Bifurcation diagrams. (a) A two-parameter stability diagram in $K_2-d_{\rm th}$ space depicting the different synchronisation transition regimes. Blue, gray, cyan, red, light green, magenta and dark gray present the regimes for continuous, tiered, classical explosive, forward and backward double jump, forward double jump, explosive with stable middle states and backward double jump synchronisation transitions, respectively. (b)--(i) $r_1$ as a function of $K_1$ for different parameter-pair values: $(K_2,d_{\rm th})=(100,0.02), (80,0.12),~(100,0.07),~(145,0.1),~(100,0.1),~(100,0.12),~(120,0.1)$ and $(120,0.08)$ respectively. Black solid and dashed lines indicate stable and unstable solutions of the self-consistent equations, respectively. Numerically simulated data points for forward and backward directions are represented by filled coloured circles connected by solid and dashed lines, respectively. The colours of the numerical data points are the same as the regimes in (a). All other parameters were fixed ($a=0$ and $b=10$).}
    \label{UD_two_parameter}
\end{figure*}

Further calculations using the Ott--Antonsen ansatz \cite{ott2008low, adhikari2023synchronization} give the reduced order model (see the Supplemental Material, Sec.\ VI):
\begin{eqnarray}
\dot{\alpha}+i\omega_0\alpha+\Delta\alpha-\frac{K_1r_1^a}{2}\sum_{k'}N(k')p^{(2)}(k,k')[\alpha(k')e^{i\beta} - &&\bar{\alpha}(k')\alpha^2(k)e^{-i\beta}]-\frac{K_2r_1^b}{2}\sum_{k',k''}N(k')N(k'') p^{(3)}(k,k',k'')    \nonumber\\
    &&[\alpha^2(k')\bar{\alpha}(k'')e^{i\beta}- \bar{\alpha}^2(k')\alpha(k'')\alpha^2(k)e^{-i\beta}]=0   \label{reduced_model}
\end{eqnarray}
The global order parameters can be calculated as
\begin{eqnarray}
    &&z_1(t)=\frac{1}{N\langle k^{(1)}\rangle}\sum_{k,k'}N(k)N(k')p^{(2)}(k,k')\bar{\alpha}(\omega_0-i\Delta,k',t) \label{eq_r1_gen}\\
    &&z_2(t)= \frac{1}{2N\langle k^{(2)}\rangle}\sum_{k,k',k''}N(k)N(k')N(k'')p^{(3)}(k,k',k'') \bar{\alpha}^2(\omega_0-i\Delta,k',t) \alpha(\omega_0 -i\Delta, k'',t),  \label{eq_r2_gen}
\end{eqnarray}
where $p^{(2)},~p^{(3)}$ are the link and triangle connection probabilities, respectively, $k,~k'$ are node degrees and $\alpha$ is the coefficient in the Fourier series of the density function. 
Solving Eqs. (\ref{reduced_model}), (\ref{eq_r1_gen}) and (\ref{eq_r2_gen}) gives the synchronisation profiles for different network topologies. 
%{ In sec.(\ref{Sec_3}) and sec.(\ref{sec_4}) we have taken two different kinds of networks and shown that this analytical procedure works well to describe the synchronization phenomena. }   
Now, we will examine how the global order parameter $r_1$ is affected by the adaptive parameters ($a,{~ \rm and}~ b~$), adaptation, and the higher-order coupling $K_2$, beginning with degree-correlated networks. 
\par
{\it Degree-Correlated Networks.} 
%\label{Sec_DC}
In this section, we examine the case where the network is generated based on prescribed node degrees. For this purpose, we take a degree sequence $\{k_1,k_2,\dots,k_N\}$ and create links with probability $p^2(k,k')=\frac{kk'}{N}$ and triangles with probability $p^3(k,k',k'')=\frac{2kk'k''}{(N\langle k\rangle)^2}$\cite{landry2020effect}. Here we have chosen $\langle k^{(1)}\rangle=\langle k^{(2)}\rangle$.
Using the same theoretical analysis, we can elucidate the behaviour of the generated network under order-parameter adaptation (see the Supplemental Material, Sec.VII for detailed calculations). Inserting the probabilities $p^{(2)}$ and $p^{(3)}$ into Eq. (\ref{reduced_model}) gives the self-consistent equations in terms of $U_1$ and $U_2$:
\begin{eqnarray}
    U_1=\sum_{k}\frac{kN(k)\alpha(k,U_1,U_2)}{N\langle k \rangle}, \label{sc_U1_dc}\\
    U_2=\sum_{k}\frac{kN(k)\alpha^2(k,U_1,U_2)}{N\langle k \rangle}
    \label{sc_U2_dc}
\end{eqnarray}
%\begin{scriptsize}
where
\begin{eqnarray}
    \alpha=\frac{-1+\sqrt{1+(K_1r_1^a k U_1+2K_2r_1^b k U_1U_2)^2\cos^2\beta}}{(K_1r_1^a k U_1+2K_2r_1^b k U_1U_2)\cos\beta}
    \label{sc_alpha}
\end{eqnarray}
and $r_1(t) = U_1$.
As a result, we can solve the self-consistent Eqs.\ (\ref{sc_U1_dc}) and (\ref{sc_U2_dc}) to find the order parameter values. Moreover, the relationship between $U_1$ and $U_2$ is demonstrated by $U_2\sim U_1^2$. 
Finally, the critical coupling strength can be expressed as 
\begin{eqnarray}
    K_1^c=\frac{2\langle k\rangle}{\langle k^2\rangle \cos\beta}.
    \label{onset}
\end{eqnarray}
Because $\beta$ was not investigated in this study, the first and second moments of the degree are entirely responsible for the onset of synchronisation.
%Also putting the value of $K_1$ as $K_1^c$ in (\ref{eq_a}) we see that the bifurcation is supercritical for $K_2<K_2^c$ and subcritical for $K_2>K_2^c$ where $K_2^c=\frac{\langle k^4\rangle \langle k\rangle r_1^a}{2\langle k^3\rangle \langle k^2\rangle r_1^b}K_1^c$.  \\
To confirm these analytical results, we built a network of size $N=5000$
using the recommended procedure, wherein the degree sequence is drawn from a uniform distribution. A degree-correlated power law distribution is analysed in the Supplemental Material, Sec.VII (B). 
%\subsection{Uniform Distribution}
% \end{widetext}
% \begin{figure}[h!]
%     \centering
%     \includegraphics[width=15cm]{K1_vs_r_K2_0p1_a0_b10_lag0_K2_adaptation.eps}
%     \caption{Degree Correlated Network: Uniform Degree Distribution. Partial adaptation in triadic coupling $K_2$, with $a=0,~b=10,~\beta=0$.}
%     \label{fig1}
% \end{figure}
%\begin{figure}[h!]
 %   \centering
%    \includegraphics[width=10cm]{uni_K1_vs_r_K2_0p1_a0_b10_lag0p1_adaptK2.eps}
 %   \caption{Degree Correlated Network: Uniform Degree Distribution. Partial adaptation in triadic coupling $K_2$, with $a=0,~b=10,~\beta=0.1$.}
  %  \label{fig1}
%\end{figure}
\par{\it Uniform distributions.} First, we take a degree sequence drawn randomly from a uniform distribution $\{75,76,\dots,145\}$. 
Here, the average degrees 
$\langle k \rangle$ and $\langle k^2 \rangle$ are in the order of $10^2 ~(109)$ and $10^4 ~ (12372)$ respectively. The phase-lag parameter ($\beta$) is fixed at zero. The partial adaptive technique is primarily utilised in the higher-order interaction term. We select nodes with degrees greater or equal to a threshold value ($d_{\rm th}$). The global order parameter 
$r_{1}$ 
is used in conjunction with the HOI coupling of those nodes (we are using 
$r_{1}^b$ to test the effect of $b$).
Figure \ref{UD_adaptK2_diff_percetage} shows how $d_{\rm th}$ affects the synchronisation order parameter $r_1$. The HOI coupling $K_2$ is fixed at 0.1 and the other parameters remain constant at $a=0$, $b=10$ and $\beta=0$ in this case. Note that the pairwise component is not adapted when $a=0$. The combined roles of $a$ and $b$ are discussed in the Supplemental Material, Sec.VII. The Euler method can be used to integrate Eq.~(\ref{model_network}) for numerical simulation and solve the self-consistent Eqs. (\ref{sc_U1_dc}) and (\ref{sc_U2_dc}) to obtain analytical $r_1$ values.
Using a Lorentzian distribution with a mean of $0$ and half width of $1$, the natural frequencies can be extracted.
The initial phases of the oscillators are drawn from a uniform distribution $[-\pi, \pi]$.
After sufficiently removing the transient parts, we can calculate stationary values of the order parameter 
$r_1$ in both the forward and backward directions. Forward simulation begins with the incoherent state at $K_1=-0.1$ and advances in small increments $K_1$ of $\Delta k=0.0005$ until the strong synchronised state ($K_1=5$) is reached. The final phase values of each previous solution are used to integrate the system as the initial conditions at each coupling strength (\ref{model_network}). Conversely, the backward simulation starts from the synchronised state ($K_1=5$) and decreases with the same step size as the forward simulation until the incoherent state is reached. As with the forward simulation, the final phase values of the prior solution are used as the starting point for each iteration.   
%Figure \ref{UD_adaptK2_diff_percetage} depicts the evolution of $r_1$ as a function of $K_1$ the presence of partial order parameter ($r_1$) adaptation with the triadic coupling ($K_2$) only.
 Figure \ref{UD_adaptK2_diff_percetage}(a) shows the transition of $r_1$ as a function of $K_1$ without any adaptation. The path to transition is clearly explosive in this case, which is consistent with previous studies \cite{skardal2020higher,dutta2023impact}.
 In Fig.~\ref{UD_adaptK2_diff_percetage}(b) $r_1$ was adapted to the nodes of degrees greater than or equal to $100$ ($d_{\rm th}=100$). Here, a new type of synchronisation transition scenario is observed. A weak synchronised state facilitates the transition from the incoherent to synchronised state. The solutions to the self-consistent equations show that a stable state emerges between two unstable states. As a result, the system makes two abrupt jumps, one forward and one backward. A decrease in the threshold value (implying more adapted nodes) increases the length of the generated stable state, reducing the width of both hystereses (Fig.\ \ref{UD_adaptK2_diff_percetage}(c)). 
 These are the key findings of this study. A carefully chosen subset of adapted nodes in a hypergraph can result in a double hysteresis loop. As more nodes are adapted, the network will eventually follow a classical route, that is, a continuous route to synchronisation (Fig.\ \ref{UD_adaptK2_diff_percetage}(d), with all nodes adapted).
% {\color{red} 
 %Comment:  Can we provide something more? Also discuss the critical onset. What is the average degree? What is the value of $\langle k^2 \rangle$?  }
Thus,  in the case of partial adaptation
$d_{\rm th}=100$ and $92$, 
we observe a double hysteresis loop along the transition route. The analytical derivation of the synchronisation onset (Eq.~(\ref{onset})) indicates that it is only affected by $\langle k \rangle$ and $\langle k^2 \rangle$. 
Other parameters cause the system to exhibit nonlinear effects. The 
$K_1^c$ value of the generated network is $0.0177$ for all transitions, which has been numerically verified.
In the Supplemental Material, Sec.VII(B), we explore the role of the degree threshold value ($d_{\rm th}$) and triadic coupling $K_2$ for a network with a power-law degree distribution. The same phenomenon is also observed in the degree-uncorrelated (random) network discussed in the Supplemental Material, Sec.VIII.
\par  One may wonder whether such an adaptation technique can cause other types of synchronisation transitions. For instance, what values of $d_{\rm th}$ and $K_2$ are suitable for generating tiered synchronisation \cite{rajwani2023tiered,dutta2024transition} For a given $K_2,b$ and $a$, how does the behavior of $r_1$ change when $d_{\rm th}$ is gradually varied from small (full adaptation) to the highest value (no adaptation)? 
\begin{figure}[h!]
   \centering
   \includegraphics[width=9cm]{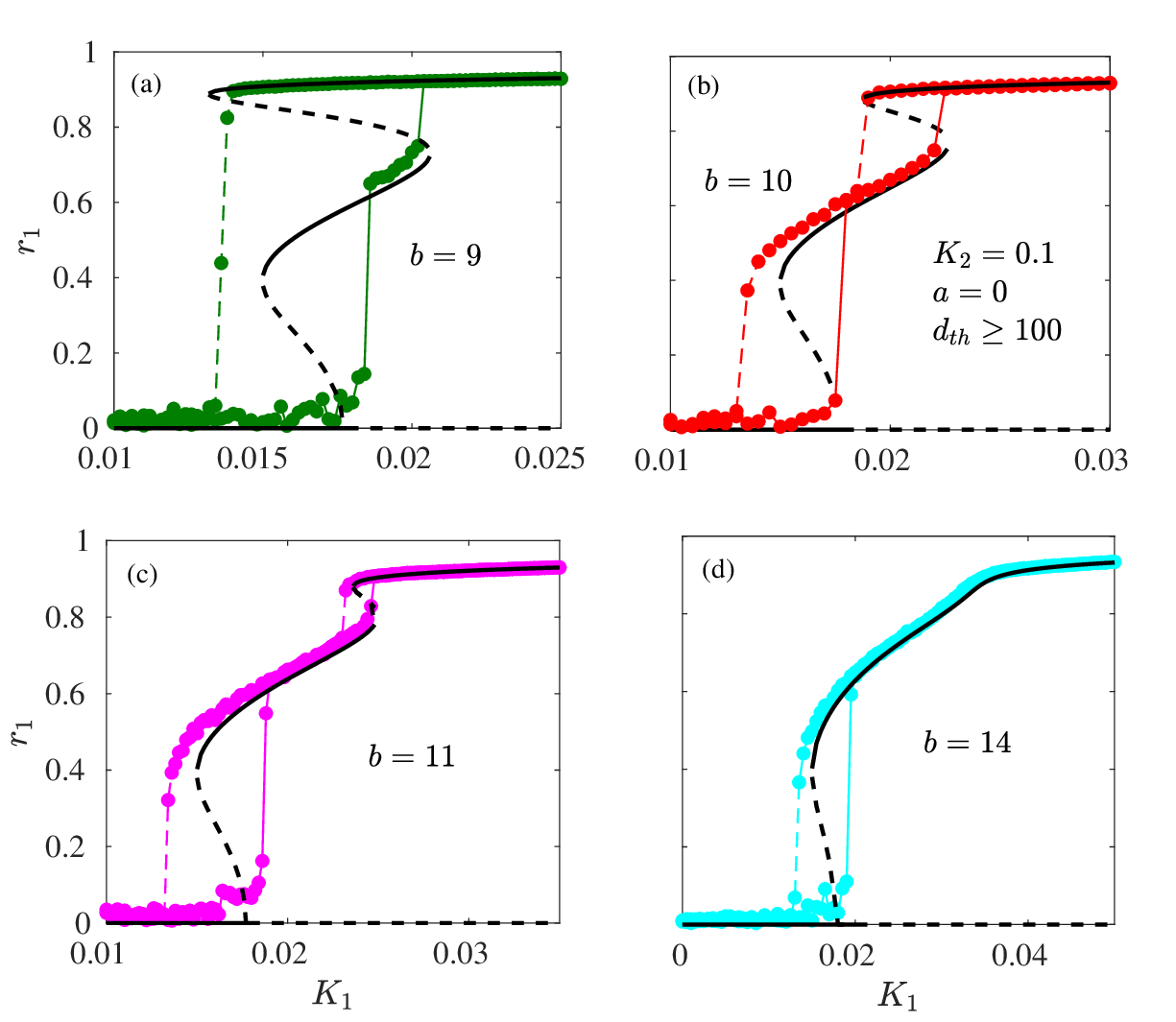}
   \caption{Synchronisation profiles showing $r_1$ as a function of $K_1$ for (a) $b=9$, (b) $b=10$, (c) $b=11$ and (d) $b=14$. Black solid and dashed lines indicate stable and unstable solutions of the self-consistent equations, respectively. Coloured (Green, red, magenta and cyan) full circles joined with solid and dashed lines represent numerically simulated data points for the forward and backward directions, respectively. All other parameters are fixed ($a=0,~d_{th}=100$ and $K_2=0.1$). }
    \label{UD_adaptK2_diff_b}
\end{figure}
%For better understanding the transition scenarios in the whole parameter space we plot the two parameter diagram Fig.\ref{UD_two_parameter} corresponding to Fig.(\ref{UD_adaptK2_diff_percetage}). 
This scenario is illustrated in Fig.\ \ref{UD_two_parameter}, with the combined effects of $d_{\rm th}$ and $K_2$ is shown.
We separated the regimes based on different synchronisation transition routes obtained from the self-consistent equations. The regimes consist of the following routes to synchronisation: Regime I (shaded in blue), represents the continuous route. Because the HOI strength is weak and a large number of nodes are adapted (bottom-left corner), pairwise coupling is dominant. This is further confirmed by Fig.\ \ref{UD_two_parameter}(b). Regime II (light gray), represents the tiered synchronisation occurring at a slightly higher HOI strength. A small, unstable branch can be observed in the upper part of the order parameter (Fig.\ \ref{UD_two_parameter}(c)). Regimes V–VII are non-trivial, and in all these cases a stable branch appears in the middle 
($r_1 \sim 0.6$). The upper ($r_1 \sim 1$) and lower stable branches ($r_1 \sim 0.0$) are connected to the middle branch by two unstable paths. In Regime V (green), the system twice jumps to synchronisation in the forward direction but in the backward direction it jumps directly to the incoherent state, whereas in Regime VII (dark gray), the system jumps twice only in backward direction. These are further confirmed by Fig.\ \ref{UD_two_parameter}(g) and (i), respectively. In the intermediate Regime VI, the system follows explosive routes (Fig.\ \ref{UD_two_parameter}(h)), though there are stable states between the forward and backward transition points. A gradual and adaptive initial phase will not reveal the middle regime; however, a proper choice of initial condition will lead the system to the stable states. In the small, narrow Regime IV (red), the system exhbits double jumps to synchronisation in both the forward and backward directions. This is further confirmed in Fig.\ \ref{UD_two_parameter}(f).
For a lower $K_2$, an increasing $d_{\rm th}$ (fewer adaptive nodes) leads to explosive synchronisation (Regime III, cyan, Fig.\ \ref{UD_two_parameter}(d) and (e)). A trade-off between adaptive nodes and the HOI strength makes the system switch from one regime to another. These two parameters contribute to the creation, annihilation and change of orientation in the middle stable branch. Four points chosen from the black horizontal line in Fig. \ref{UD_two_parameter}(a) are illustrated in Fig.\ \ref{UD_adaptK2_diff_percetage}.
%For particular 
%$K_2$, 
%if we move horizontally, we can see the effect of  
%$d_{th}$.  Now to see the effect of %$K_2$, we fix the 
%$d_{th}$ 
%value to 
%$100$ 
%and move vertically. 
%, with the increase in $K_2$
%first creates a hysteresis at the lower part of the route (Fig.\ref{UD_two_parameter}(d)), leads the system to experience explosive synchronization. Then increase in $K_2$ creates another hysteresis in the upper part of the route also (Fig.\ref{UD_two_parameter}(e)), leads to double jump in both the direction. Further increase in $K_2$ moves the backward point in the left side which leads the system to directly jump to incoherent state (Fig.\ref{UD_two_parameter}(g)). Thus combined effect of $d_{th}$ and $K_2$ generates different synchronization routes.  
\par
Next, we will explore the effect of other parameters, such as 
$a$ and $b$. 
To do this, we shall use values of $b$ as $9,~10,~11$ and $14$ to compare the transition routes reported in Fig.\ \ref{UD_adaptK2_diff_b}. The other parameters remain fixed at 
$d_{\rm th}=100$, $K_2=0.1$ and $a=0$. 
For $b=9$, the upper hysteresis broadened and the lower hysteresis remained similar to $b=10$, 
whereas for $b=11$ the width of the upper hysteresis shrink and for 
$b=14$ it vanished. 
However, the lower hysteresis remained constant. Therefore, the exponent $b$ affects only the upper part of the hysteresis. The numerical data also support our analysis. 
%{\color{red} Really interesting observation. But why is it  happening? Can we map with known bifurcation phenomena? Why does it $b$ affect so much?}
\par
\begin{figure}[h!]
    \centering
    \includegraphics[width=9cm]{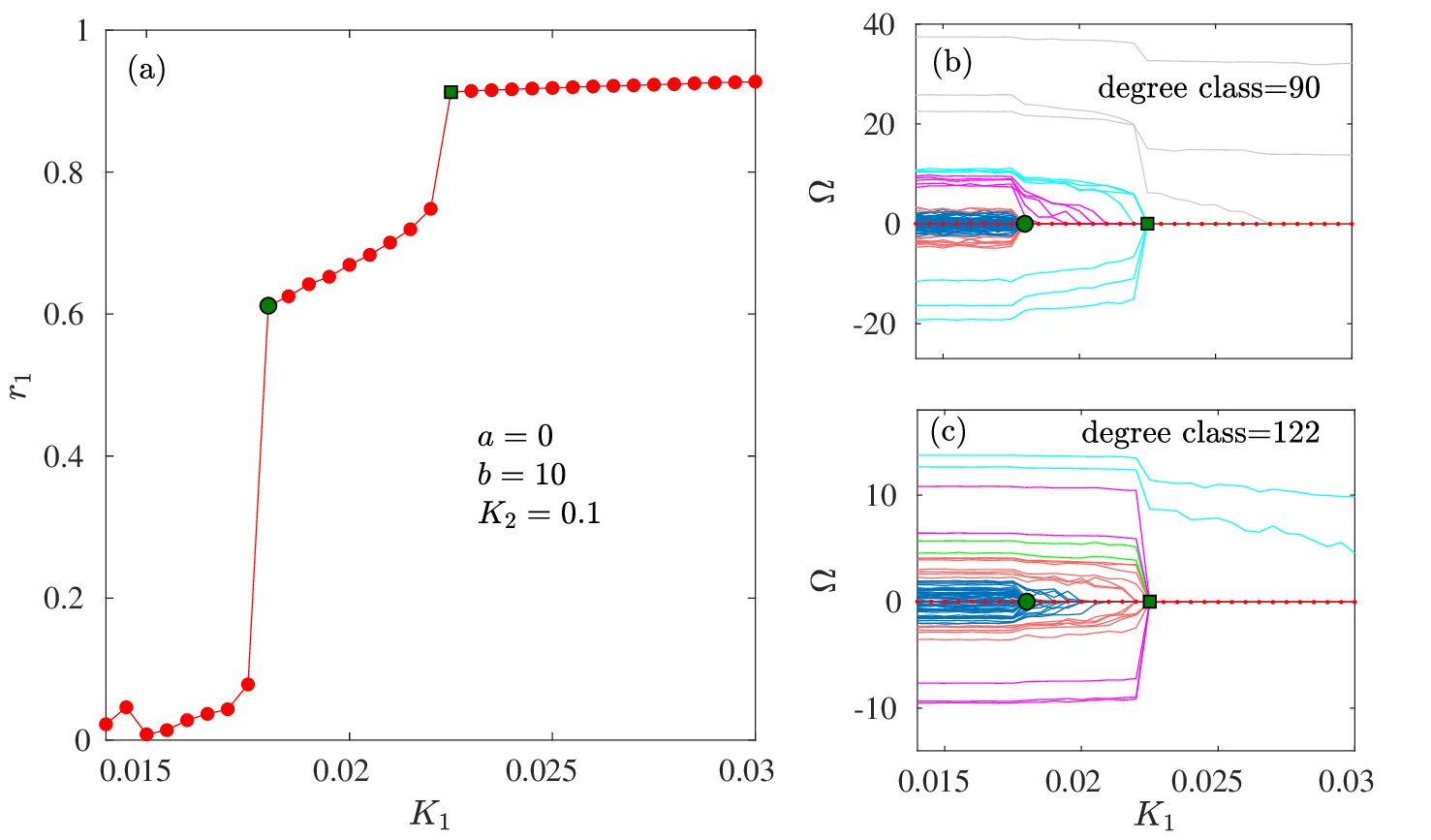}
    \caption{Synchronisation transition frequency evolution. (a) $r_1$ as a function of $K_1$ for a double jump in the forward simulation. Two transition points are indicated by a full green circle and square. Mean frequency as a function of $K_1$ for (b) $90$ (not adapted) and (c) $122$ (adapted) degree classes. The blue, light-red, light-green, magenta, cyan and gray colours represent nodes with natural frequencies in the ranges $|\omega_i|<2,~ (2,4),~(4,6),~ (6,10),~ (10,20)$ and $>20$ respectively. The dotted red line represents the mean frequency of the network.}
    \label{ER_frequency}
\end{figure}

 Finally, we will discuss the relevant mechanisms of double explosive transition. Because of partial adaptation, a new stable branch emerges in the middle of the synchronisation transition. This raises the question of which nodes are involved in the synchronisation transition on that particular branch. 
Thus, to visualise the node behaviour, we computed the effective frequency, as given by
\begin{eqnarray}
    \Omega_i=\frac{1}{T}\int_t^{t+T}\dot{\theta}_i(\tau)d\tau,
\end{eqnarray}
where $T$ is the total simulation time. 
Figure \ref{ER_frequency} depicts the synchronisation transition and corresponding effective frequencies at parameter values 
$K_2=0.1,~ a=0,~ b=10,$ and $d_{th}=95$.
We adapted the order parameter $r_1$ to the nodes with degrees greater than a threshold value ($d_{\rm th}$). Thus,  we  have shown the evolution of $\Omega_i$ for two degree classes---one adapted ($122$) and another non-adapted ($90$).
The frequencies in Fig. \ref{ER_frequency} are coloured according to the range of their natural frequencies: blue, red, light green, magenta, cyan and gray for $|\omega_i|<2,~ (2,4),~(4,6),~ (6,10),~(10,20)$ and $>20$ respectively.
The figure clearly demonstrates that, at the two critical points, the frequencies jump abruptly to the average frequency. Figure \ref{ER_frequency}(b) shows that the non-adapted nodes join the synchronised cluster at the first (second) transition point in the range $|\omega_i|<6$ ($|\omega_i|<20$) approximately. The adapted nodes whose frequencies are very close to $0$ join the synchronised cluster at the first transition point, and adapted nodes with $|\omega_i|<10$ join at the second transition point (Fig. \ \ref{ER_frequency}(c)). Subsequently, the remaining nodes with frequencies $|\omega_i|>10$ join one by one as the coupling strength increases. This occurs due to the order parameter adaptation, which reduces the effective coupling strength. Similarly, we computed the frequency $\Omega$ for other synchronisation transitions (see the Supplementary Material, Sec.IX). This clarifies the basic mechanisms behind the double jump in the transition paths.       
\section{Discussion}
In this study, a general principle for realising a double explosive transition in forward, backward or a combination of both directions was proposed. The use of triadic interactions and the partial adaptation of the global order parameter operating on the triadic coupling were found to be necessary for the creation of such a phase transition. We tested this technique with correlated uniform and power-law graphs. We also observed this phenomenon in uncorrelated random hypergraphs. The critical onset of synchronisation was also derived. The partial adaptation and the exponents in the adaptation order parameter tune and control the widths of the double explosive transition. In future studies, we will investigate triple explosive transitions in directed hypergraphs.
\section{ACKNOWLEDGMENTS}
S.D. acknowledges the support from DST, India under the
INSPIRE program (Code No. IF190605). L.M. gratefully acknowledges the support of the ``Hundred Talents" program of the University of Electronic Science and Technology of China, of the ``Outstanding Young Talents Program (Overseas)" program of the National Natural Science Foundation of China, and of the talent programs of the Sichuan province and Chengdu municipality.
%\newpage
\bibliographystyle{apsrev4-1}
%\bibliography{Pitu_master_bib}
\bibliography{References}
%\newpage

\newpage

\title{Supplemental Material: A double explosive Kuramoto transition in hypergraphs}
\author{Sangita Dutta$^1$}
\email{sangitaduttaprl@gmail.com}
\author{Prosenjit Kundu$^2$}
\author{Pitambar Khanra$^3$}
\author{Ludovico Minati$^{4,5}$}
\author{Stefano Boccaletti$^{6,7,8}$}
\author{Pinaki Pal$^1$}
\email{ppal.maths@nitdgp.ac.in}
\author{Chittaranjan Hens$^9$}
\affiliation{$^1$Department of Mathematics, National Institute of Technology, Durgapur~713209, India}
\affiliation{$^2$Dhirubhai Ambani Institute of Information and Communication Technology, Gandhinagar, Gujarat, 382007, India}
\affiliation{$^3$Department of Chemical and Biological Engineering, The State University of New York at Buffalo, Buffalo, New York 14260, USA}
\affiliation{$^4$School of Life Science and Technology, University of Electronic Science and Technology of China, 611731 Chengdu, China}
\affiliation{$^5$Center for Mind/Brain Sciences (CIMeC), University of Trento, 38123 Trento, Italy}
\affiliation{$^6$Research Institute of Interdisciplinary Intelligent Science, Ningbo University of Technology, 315104 China}
\affiliation{$^7$Sino-Europe Complexity Science Center, North University of China, 030051, Taiyuan, China}
\affiliation{$^8$CNR - Institute of Complex Systems, Via Madonna del Piano 10, I-50019, Sesto Fiorentino, Italy}
\affiliation{$^9$Center for Computational Natural Science and Bioinformatics, International Institute of Informational Technology, Gachibowli, Hyderabad 500032, India}

\maketitle

\section{Supplemental Material: Introduction}
In our main text we explore different synchronization transition scenarios caused by partial order-parameter adaptation in complex networks. In this supplement, we provide detailed analytic derivations of the self-consistent equations and validate them with different parameter values for different kinds of network. Following the same procedure of partial order-parameter adaptation in the triadic coupling $K_2$, we describe the case when $r_1$ is partially adapted to the pairwise coupling $K_1$. We also prove that by rescaling the coupling strengths, we may obtain results for an all-to-all connected network. Finally, we will present the frequency evolutions for different synchronization transitions. 

\section{Hypergraphs}
%In this section we will focus on hypergraphs. 
Mathematically, hypergraph models are able to encode higher-order interactions in complex systems, beyond pairwise interactions. They are formed by a pair of node and hyperlink sets, where the hyperlinks describe the group interactions among the nodes. There are several frameworks for constructing hypergraphs \cite{courtney2016generalized,landry2020effect}. Here, we chose a method where for a specified vector of hyperdegrees $[k_1, k_2,\dots,k_N]$, a hyperlink $(i_1,i_2,i_3,\dots,i_m)$ is created with probability $p^{(m)}(k_{i_1},k_{i_2},\dots,k_{i_m})$. In this supplement we consider higher-order interactions up to triangular interactions. Therefore, we have hyperlinks of size $2$ and $3$, i.e. links and triangles, respectively.

\section{Analytic Derivation}
The equation for the system is given as    
\begin{eqnarray}
\label{model_net_supp}
\dot{\theta_i}=\omega_i+K_1 r_1^a \sum_{j=1}^{N} A_{ij} \sin(\theta_j-\theta_i-\beta)+K_2r_1^b \sum_{j=1}^N \sum_{k=1}^N B_{ijk} \sin(2\theta_j-\theta_k-\theta_i-\beta), \hspace*{0.5cm} i=1,2,\dots,N. 
\end{eqnarray}
Here, our first concern is to reduce the dimensionality of the model (\ref{model_net_supp}) so that we might easily track the system analytically. Therefore, we substitute the local order parameters (see the main text) in Eq.(\ref{model_net_supp}), yielding  
\begin{equation}
\label{reduction_model_net}
\dot{\theta_i}=\omega_i+\frac{1}{2i}\left[e^{-i(\theta_i+\beta)}H_i-e^{i(\theta_i+\beta)}\bar{H_i}\right],
\end{equation}
where $H_i=K_1r_1^a R_i^1+K_2 r_1^b R_i^2$. \\
To further simplify calculations, we can assume that nodes with the same hyperdegree $k$ are statistically equivalent \cite{adhikari2023synchronization}. Thus, we rewrite the local order parameters as  
\begin{eqnarray*}
&&R_i^1\rightarrow R^1(k_i,t), \\
&&R_i^2\rightarrow R^2(k_i,t).
\end{eqnarray*}
These order parameters quantify the macroscopic dynamics of the network. 
In the continuum limit, $R^1(k)$ and $R^2(k)$ can be expressed as  
\begin{widetext}
  \begin{eqnarray}
    R^1(k)&=&\sum_{k'}N(k')p^{(2)}(k,k')\int\int f(\theta',\omega',k',t)e^{i\theta'}d\theta' d\omega', \label{eq_R1}\\
    R^2(k)&=&\sum_{k',k''}N(k')N(k'')p^{(3)}(k,k',k'')\int\int\int\int f(\theta',\omega',\theta'',\omega'',k',k'',t)e^{2i\theta'}e^{-i\theta''}d\theta' d\omega' d\theta'' d\omega'' \nonumber \\
    &\approx& \sum_{k',k''}N(k')N(k'')p^{(3)}(k,k',k'')\int\int f(\theta',\omega',k',t) e^{2i\theta'}d\theta' d\omega' \int\int f(\theta'',\omega'',k'',t) e^{-i\theta''} d\theta'' d\omega''.    \label{eq_R2}
\end{eqnarray}  
\end{widetext}
where 
%$f(\theta, \omega, k, t)$ is the density function of the oscillators such that 
$f(\theta', \omega', k', t)d\theta' d\omega'$ represents the density of oscillators with a natural frequency lying in the range $[\omega', \omega'+d\omega']$ and phases lying in $[\theta'+d\theta']$, $p^{(2)}(k,k')$ is the probability of creating a link by connecting nodes with hyperdegrees of $k$ and $k'$, $f(\theta',\omega',\theta'',\omega'',k',k'',t)$ describes the joint density of two oscillators with a phase, frequency and hyperdegree of $\theta',~\omega',~k'$ and $\theta'',~\omega'',~k''$, respectively; and $p^{(3)}(k,k',k'')$ is the probability of creating a triangle connecting nodes with hyperdegrees of $k,~k'$ and $k''$. In the limit of complete incoherent and synchronized states or in the case of a higher mean hyperdegree, the following assumption can be made: 
\begin{equation*}
    f(\theta',\omega',\theta'',\omega'',k',k'',t)=f(\theta',\omega',k',t) f(\theta'',\omega'',k'',t).
\end{equation*}
This relation holds well for dense hypergraphs. Moreover, the oscillators are conserved in the system and the density function must satisfy the continuity equation
\begin{equation}
    \frac{\partial f}{\partial t}+\frac{\partial}{\partial \theta}{\left(f\left(\omega+\frac{1}{2i}\left[e^{-i(\theta+\beta)}H-e^{i(\theta+\beta)}\bar{H}\right]\right)\right)}=0.
    \label{continuity_eq2_supp}
\end{equation}
% Thereafter expanding the density function $f$ in furrier series same as Eq.(\ref{ott-FS}) and applying the Ott-Antonsen assumption, 
Subsequently, we expand the density function in the Fourier series as
\begin{eqnarray*}
    f(\theta,\omega,k,t)=\frac{g(\omega)}{2\pi}\left[1+\sum_{n=1}^\infty \left[f_n e^{in\theta}+\Bar{f}_ne^{-in\theta}\right]\right]. 
\end{eqnarray*}
Next, we follow the renowned Ott--Antonsen ansatz \cite{ott2008low}, which states that the coefficients $f_n$ can assume the form $f_n=\alpha^n$ for some analytic function $\alpha$ with $|\alpha|\leq 1$. 
Applying this ansatz, the continuity Eq. (\ref{continuity_eq2_supp}) reduces to 
\begin{equation}
\label{eq_alpha_net}
\dot{\alpha}+i\alpha\omega-\frac{1}{2}\left[\bar{H}e^{i\beta}-H\alpha^2e^{-i\beta}\right]=0.
\end{equation}
Substituting the Fourier expansion of $f$ into Eqs. (\ref{eq_R1}) and (\ref{eq_R2}) yields
\begin{widetext}
    \begin{eqnarray}
        R^1(k)&=&\sum_{k'}N(k')p^{(2)}(k,k')\int\int\frac{g(\omega')}{2\pi}\left[1+\sum_{n=1}^\infty \left[\alpha^n e^{in\theta'}+\Bar{\alpha}^ne^{-in\theta'}\right]\right]e^{i\theta'}d\theta' d\omega'   \nonumber \\
        &=& \sum_{k'}N(k')p^{(2)}(k,k')\int g(\omega')\bar{\alpha}(\omega',k',t) d\omega',
    \end{eqnarray}
    \begin{eqnarray}
      R^2(k)=\sum_{k',k''}N(k')N(k'')p^{(3)}(k,k',k'')\int g(\omega')\bar{\alpha}^2(\omega',k',t) d\omega'  \int g(\omega'')\alpha(\omega'',k'',t) d\omega''.
    \end{eqnarray}
\end{widetext}
For the chosen Lorentzian frequency distribution, $g(\omega)=\frac{\Delta}{\pi[\Delta^2+(\omega-\omega_0)^2]}$, the contour integration can be performed in the lower-half $\omega$ plane. Therefore, the order parameters are evaluated as   
%\begin{widetext}
    \begin{eqnarray}
        R^1(k)=\sum_{k'}N(k')p^{(2)}(k,k')\bar{\alpha}(\omega_0-i\Delta,k',t), 
        \label{eq_R1_value}
    \end{eqnarray}
    \begin{eqnarray}
        R^2(k)=\sum_{k',k''}N(k')N(k'')p^{(3)}(k,k',k'')\bar{\alpha}^2(\omega_0-i\Delta,k',t)\alpha(\omega_0 -i\Delta, k'',t). \label{eq_R2_value}
    \end{eqnarray}
Calculating Eq. (\ref{eq_alpha_net}) at $\omega=\omega_0-i\Delta$ gives
    \begin{eqnarray*}
        \dot{\alpha}+i\alpha(\omega_0-i\Delta)-\frac{1}{2}[K_1r_1^a(\bar{R}_i^1e^{i\beta}-R_i^1\alpha^2e^{-i\beta})+K_2r_1^b(\bar{R}_i^2e^{i\beta}-R_i^2\alpha^2e^{-i\beta})]=0.
    \end{eqnarray*}
%\end{widetext}
Substituting the order parameter values from Eq. (\ref{eq_R1_value}) and Eq. (\ref{eq_R2_value}), we get
\begin{eqnarray}
    \dot{\alpha}+i\omega_0\alpha+\Delta\alpha-\frac{K_1r_1^a}{2}\sum_{k'}N(k')p^{(2)}(k,k')[\alpha(k')e^{i\beta}&&-\bar{\alpha}(k')\alpha^2(k)e^{-i\beta}]-\frac{K_2r_1^b}{2}\sum_{k',k''}N(k')N(k'') p^{(3)}(k,k',k'')    \nonumber\\
    &&[\alpha^2(k')\bar{\alpha}(k'')e^{i\beta}-\bar{\alpha}^2(k')\alpha(k'')\alpha^2(k)e^{-i\beta}]=0.    \label{reduced_model_supp}
\end{eqnarray}
Thus, for a given network topology we can find a reduced model that describes the original model.  

\section{Degree-correlated network}
As described in the main text, we chose the probability values and generated a synthetic network with degree--degree correlation \cite{adhikari2023synchronization}. To obtain the reduced model, we insert the values of the probabilities into Eq. (\ref{reduced_model_supp}), yielding  
\begin{eqnarray}
    \dot{\alpha}+i\omega_0\alpha+\Delta\alpha-\frac{K_1r_1^a}{2}\sum_{k'}N(k')\frac{kk'}{N\langle k \rangle}[\alpha(k')e^{i\beta}&&-\bar{\alpha}(k')\alpha^2(k)e^{-i\beta}]-\frac{K_2r_1^b}{2}\sum_{k',k''}N(k')N(k'') \frac{2kk'k''}{(N\langle k \rangle)^2}    \nonumber\\
    &&[\alpha^2(k')\bar{\alpha}(k'')e^{i\beta}-\bar{\alpha}^2(k')\alpha(k'')\alpha^2(k)e^{-i\beta}]=0.   \label{eq_alpha_dc}
\end{eqnarray}
Now we assume
\begin{eqnarray}
    U_1=\sum_{k'}\frac{k'N(k')\alpha(k')}{N\langle k \rangle} \label{U1_dc}\\
  \text{and} \hspace{0.5cm}  U_2=\sum_{k'}\frac{k'N(k')\alpha^2(k')}{N\langle k \rangle}.
    \label{U2_dc}
\end{eqnarray}
Substituting these expressions for $U_1,~U_2$ into Eq. (\ref{eq_alpha_dc}) simplifies it to 
\begin{eqnarray}
    \dot{\alpha}+i\omega_0\alpha(k)+\Delta\alpha(k)-\frac{K_1r_1^a k }{2}[U_1e^{i\beta} -\bar{U}_1\alpha^2(k)e^{-i\beta}]-K_2r_1^b k[U_2\bar{U}_1e^{i\beta}-\bar{U}_2U_1\alpha^2(k)e^{-i\beta}]=0.
\end{eqnarray}
Here, we want stationary rotating solutions of $\alpha$. Thus, we use the polar forms $\alpha(k,t)=\alpha e^{i\omega_1 t}$, $U_1(t)=U_1e^{i\omega_1 t}$ and $U_2(t)=U_2e^{2i\omega_1 t}$. Then, taking $\Delta=1$ and separating the real and imaginary parts we get 
\begin{eqnarray}
    \alpha-(\frac{K_1}{2}r_1^a k U_1+K_2r_1^b k U_1U_2)(1-\alpha^2)\cos\beta=0,
    \label{real_eq_alpha_dc}
\end{eqnarray}
\begin{eqnarray}
    \alpha\omega_1 =-\alpha\omega_0+(\frac{K_1}{2}r_1^a k U_1+K_2r_1^b k U_1U_2)(1+\alpha^2) \sin\beta.
\end{eqnarray} 
The solution of Eq. (\ref{real_eq_alpha_dc}) in a form with $\alpha$ can be combined with Eqs. (\ref{U1_dc}) and (\ref{U2_dc}) to form the self-consistent equations in terms of $U_1$ and $U_2$, expressed as
\begin{eqnarray}
U_1=\sum_{k}\frac{kN(k)\alpha(k,U_1,U_2)}{N\langle k \rangle}, \label{sc_U1_dc_supp}\\
\text{and} \hspace{0.5cm}  
    U_2=\sum_{k}\frac{kN(k)\alpha^2(k,U_1,U_2)}{N\langle k \rangle}
    \label{sc_U2_dc_supp}
\end{eqnarray}
%\begin{scriptsize}
where
\begin{eqnarray}
    \alpha=\frac{-1+\sqrt{1+(K_1r_1^a k U_1+2K_2r_1^b k U_1U_2)^2\cos^2\beta}}{(K_1r_1^a k U_1+2K_2r_1^b k U_1U_2)\cos\beta}.
    \label{sc_alpha_dc_supp}
\end{eqnarray}
%\end{scriptsize}
To find a complete description of the synchronisation profiles, we now consider the definition of the order parameter values (Eqs. (\ref{eq_R1_value}) and (\ref{eq_R2_value})) and substitute the probability values $p^{(2)}$ and $p^{(3)}$, yielding 
\begin{eqnarray*}
    z_1(t)&=&\frac{1}{N\langle k \rangle}\sum_{k,k'}N(k)N(k')\frac{kk'}{N\langle k \rangle}\bar{\alpha}(\omega_0-i\Delta,k',t) \\
    &=&\bar{U}_1, \\
  z_2(t) &=& \bar{U}_2 U_1.
\end{eqnarray*}
Therefore, $r_1=|U_1|=U_1$ and $r_2=|U_2U_1|=U_2U_1$. This gives the relationship between $U_1$, $r_1$ and $U_2$, $r_2$, which, in combination with the self-consistent equations (\ref{sc_U1_dc_supp}), (\ref{sc_U2_dc_supp}) and (\ref{sc_alpha_dc_supp}), will provide the synchronisation profiles for different parameter values.

Next, to determine the type of bifurcation at the onset of synchronisation, we expand $\alpha$ in Eq. (\ref{sc_alpha_dc_supp}) up to the cubic terms, giving
\begin{eqnarray}
    U_1&=&\frac{1}{N\langle k\rangle}\sum_k N(k)k [\frac{1}{2}K_1r_1^akU_1\cos\beta+K_2r_1^bkU_1U_2\cos\beta-\frac{1}{8}(K_1r_1^a k U_1\cos\beta)^3]. 
    \label{eq_U1_c}
\end{eqnarray}
Simplifying this gives
\begin{eqnarray}
    U_1&=&\frac{\langle k^2\rangle}{2\langle k\rangle}K_1r_1^aU_1\cos\beta+\frac{\langle k^2\rangle}{\langle k\rangle}K_2r_1^bU_1U_2\cos\beta-\frac{\langle k^4 \rangle}{8\langle k \rangle}(K_1r_1^aU_1\cos\beta)^3,  \label{U1_expand}
\end{eqnarray}
where $\langle k^2 \rangle=\frac{1}{N}\sum_k N(k)k^2$ and $\langle k^4 \rangle=\frac{1}{N}\sum_k N(k)k^4$. Similarly, $U_2$ can be expressed as
\begin{eqnarray}
    U_2&=&\frac{\langle k^3\rangle}{4\langle k\rangle}(K_1r_1^aU_1\cos\beta)^2.
\end{eqnarray}
Because Ref. \cite{dutta2024transition} showed that in cases of non-zero $a$, $r_1=0$ always remains a stable, fixed point of the system, we shall use $a=0$ to find the critical coupling strength. Thus, we take the limit $U_1\rightarrow 0^+$ in Eq. (\ref{U1_expand}). The remaining terms provide the critical coupling strength:
\begin{eqnarray}
    K_1^c=\frac{2\langle k\rangle}{\langle k^2\rangle \cos\beta}.
\end{eqnarray}
Simplifying $U_1$ in Eq. (\ref{U1_expand}) gives
\begin{eqnarray}
    \frac{\langle k^4 \rangle}{8\langle k \rangle}(K_1\cos\beta)^3 U_1^2-
    \frac{\langle k^2\rangle \langle k^3 \rangle}{4\langle k\rangle^2}K_1^2K_2\cos^3\beta U_1^{(b+2)} 
    =\frac{K_1}{K_1^c}-1. 
\end{eqnarray}
% where 
% \begin{eqnarray}
%     A=\frac{\langle k^4 \rangle}{8\langle k \rangle}(K_1\cos\beta)^3-
%     \frac{\langle k^2\rangle \langle k^3 \rangle}{4\langle k\rangle^2}K_1^2K_2r_1^{b}\cos^3\beta
%     \label{eq_a}
% \end{eqnarray} 
Clearly, a bifurcation occurs at $K_1=K_1^c$. This bifurcation is subcritical if a solution for $K_1<K_1^c$ exists and supercritical otherwise, for different choices of parameter values.

However, in the cases of partially adapting $r_1$ to $K_1$ with the threshold value ($d_{th}$) in presence of adaptation with $K_2$, we can find the critical coupling strength for synchronisation transition by taking $U_1\rightarrow0^+$ in Eq. (\ref{eq_U1_c}), giving
\begin{eqnarray}
    K_1^c=\frac{2N\langle k\rangle}{\sum_{k_{\rm min}}^{d_{\rm th}-1} N(k) k^2 \cos\beta},
\end{eqnarray}
where $k_{\rm min}$ is the minimum node degree.

\subsection{Uniform Distribution}
\label{sec_UD_supp}
% \begin{figure}[h!]
%    \centering
%    \includegraphics[width=9cm]{K1_vs_r_a0_b10_lag0_deg_geq100_K2_adaptation.eps}
%    \caption{Degree Correlated Network: Uniform Degree Distribution. Partial adaptation in pairwise coupling $K_1$, with $a=0,~b=10,~\beta=0$.}
%     \label{UD_adaptK2_diff_K2}
% \end{figure}
% Similarly next we fix $b=10$, $d_{th}=100$ and take different values of $K_2$ in Fig.\ref{UD_adaptK2_diff_K2}. Here we see a completely different scenario compared to Fig.\ref{UD_adaptK2_diff_b}. For small triadic coupling ($K_2=0.08$) only lower hysteresis exists. As $K_2$ value is increased ($K_2=0.1$), the upper hysteresis is generated. Further increase in $K_2$ value pull the backward transition point of the upper hysteresis in the backward direction. Note that here also the lower hysteresis part is not much affected by the triadic coupling $K_2$. Therefore from the above analysis we can say that both the upper and lower hysteresis is generated as an effect of partial order parameter adaptation in the triadic coupling $K_2$, while the parameters $b$ and $K_2$ mainly control the upper hysteresis part. 

\begin{figure}[h!]
   \centering
   \includegraphics[width=9cm]{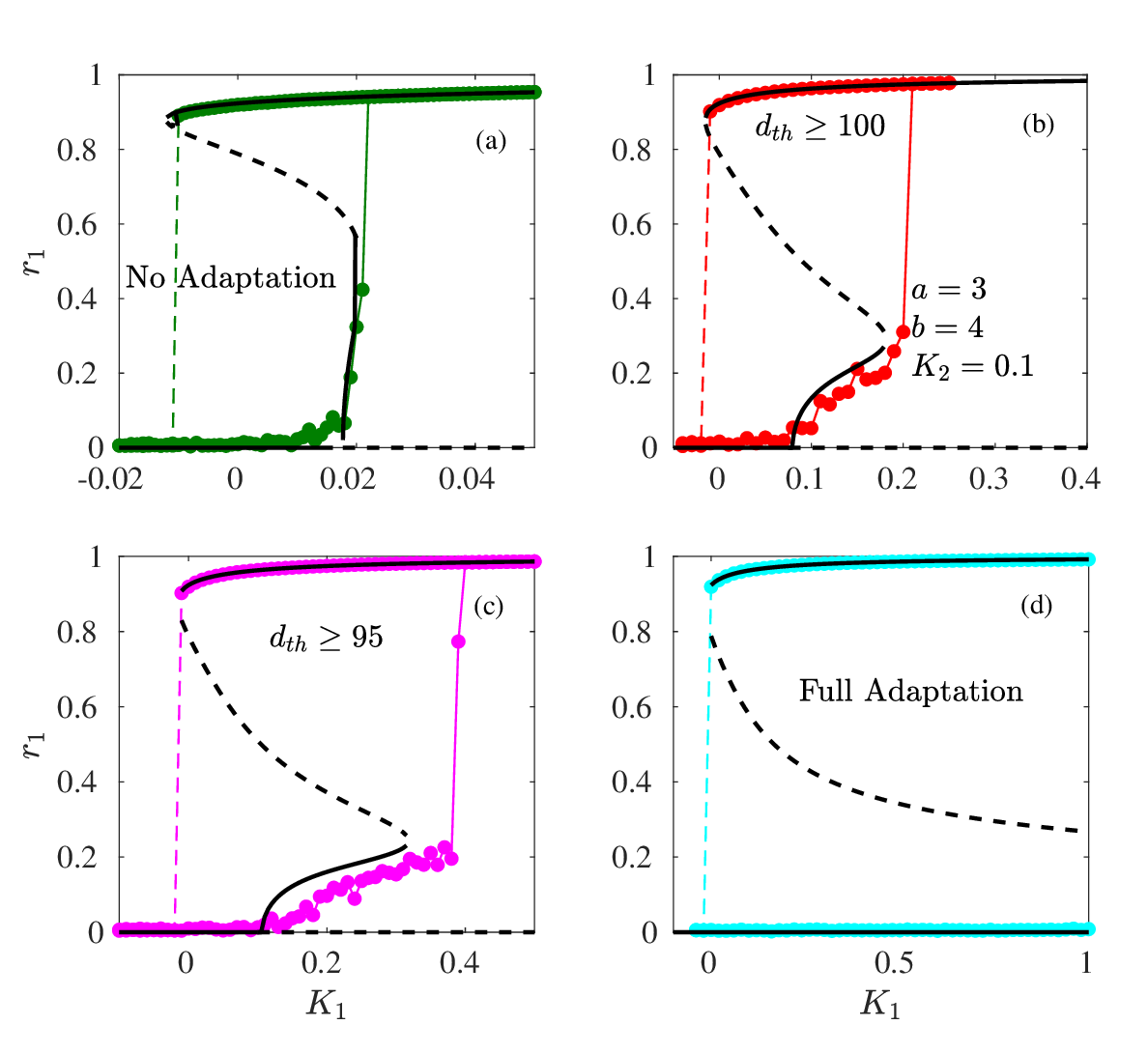}
   \caption{Synchronisation profiles for partially adapting $r_1$ to $K_1$ in a degree-correlated network with a uniform degree distribution. $r_1$ as a function of $K_1$ for (a) no adaptation, (b) $d_{\rm th}=100$, (c) $d_{th}=95$ and (d) full adaptation. Black solid and dashed lines indicate stable and unstable solutions of the self-consistent equations, respectively. Green, red, magenta and cyan full circles joined with solid and dashed lines represent numerically simulated data for the forward and backward directions, respectively. All other parameters are fixed ($a=3,~b=4$ and $K_2=0.1$).}
    \label{UD_adaptK1}
\end{figure}
In the main text, we analysed the effect of partial order-parameter adaptation with triadic coupling $K_2$ in the absence of adaptation with pairwise coupling. Our next concern is to see the effect of partially adapting $r_1$ to the pairwise coupling $K_1$. Here, we set the parameter values as $K_2=0.1,~a=3$ and$~b=4$. We adapt the global order parameter $r_1$ with pairwise coupling $K_1$ in a fraction of nodes with degrees greater than or equal to a threshold degree value $d_{\rm th}$.  Figure (\ref{UD_adaptK1}) shows the synchronisation profiles for different $d_{\rm th}$ with varying $K_1$. We see that in the case of no adaptation to $K_1$, the path to synchronisation is tiered (Fig. \ref{UD_adaptK1}(a)). For $d_{th}=100$, the forward transition point moves forward and the length of the lower stable state is increased (Fig. \ref{UD_adaptK1}), although the synchronisation level is decreased. In Fig. \ref{UD_adaptK1}(c), for $d_{th}=95$, the forward jump point moves further away from the onset. Finally, for full order parameter adaptation, the $r_1=0$ fixed point remains stable for all $K_1$ values and the transition becomes explosive. The reason behind this transformation is that increasing the number of adaptive nodes reduces the effective pairwise coupling strength $K_1$. Therefore, for the dominating nature of $K_2$, the system tends to follow explosive paths to synchronisation. Here, the numerically simulated data points match closely with the analytical ones.

\subsection{Power-Law Distribution}
\begin{figure}[h!]
   \centering
   \includegraphics[width=9cm]{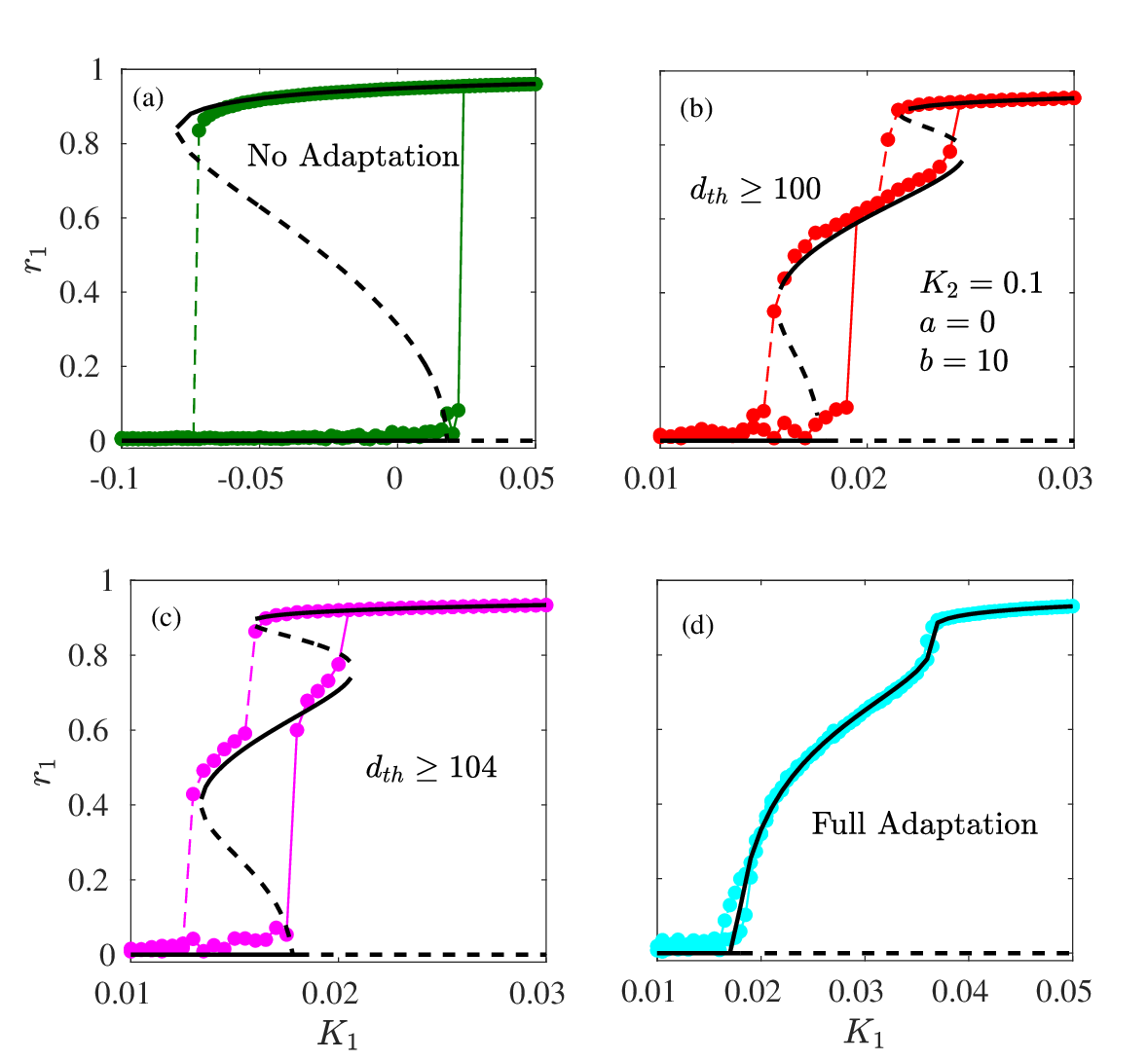}
   \caption{Synchronization profiles for adapting $r_1$ partially with $K_2$ in a degree-correlated network with a power-law degree distribution. $r_1$ as a function of $K_1$ for (a) no adaptation, (b) $d_{th}=100$, (c) $d_{th}=104$ and (d) full adaptation. Black solid and dashed lines indicate stable and unstable solutions of the self-consistent equations, respectively. Green, red, magenta and cyan full circles joined with solid and dashed lines represent numerically simulated data for the forward and backward directions, respectively. All other parameters are fixed ($a=0,~b=10$ and $K_2=0.1$).}
    \label{SF_adapt_K2}
\end{figure}

\begin{figure}[h!]
    \centering
    \includegraphics[width=9cm]{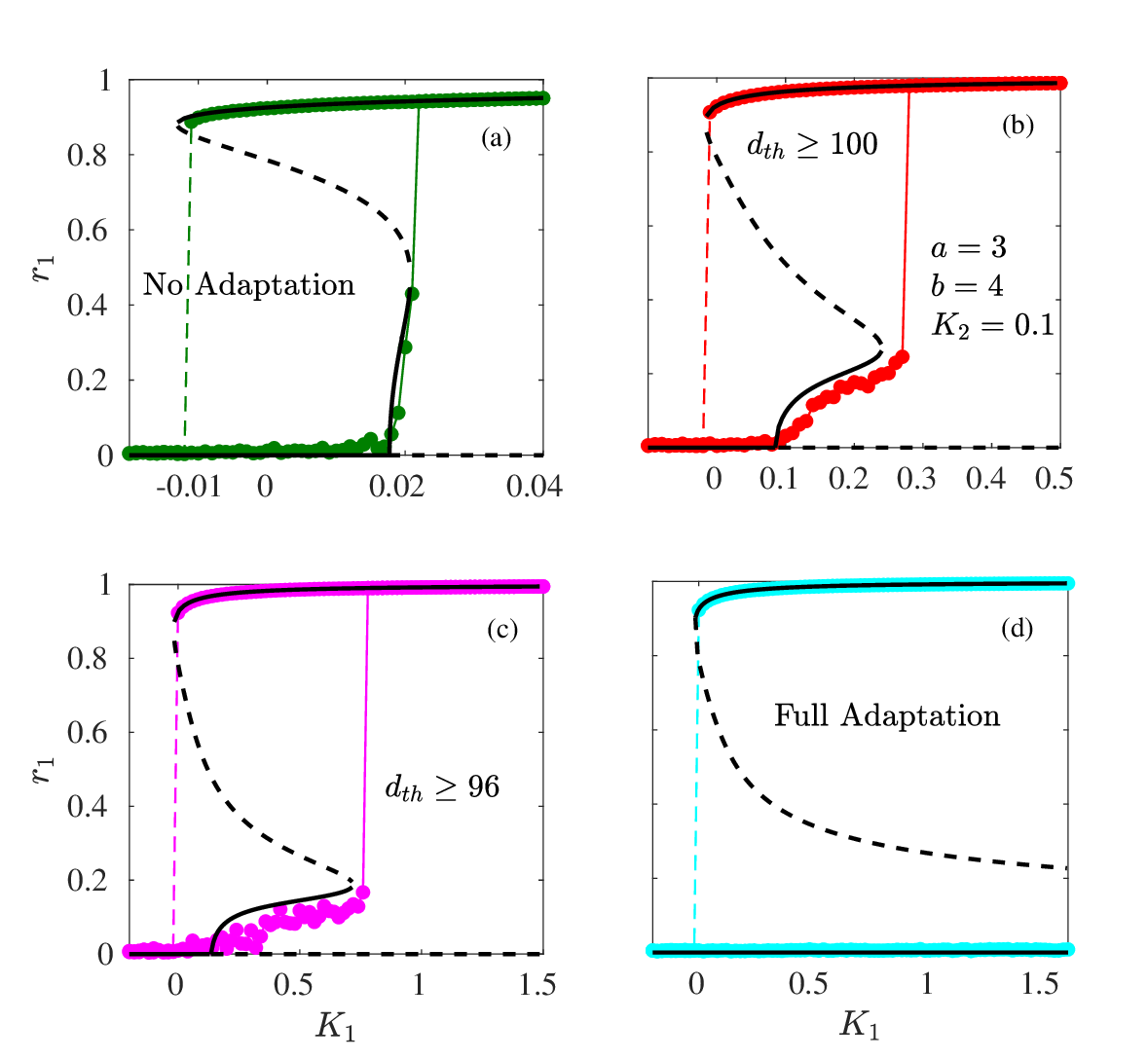}
    \caption{Synchronisation profiles for adapting $r_1$ partially to $K_1$ in a degree-correlated network with a power-law degree distribution. $r_1$ as a function of $K_1$ for (a) no adaptation, (b) $d_{th}=100$, (c) $d_{th}=96$ and (d) full adaptation. Black solid and dashed lines indicate stable and unstable solutions of the self-consistent equations, respectively. Green, red, magenta and cyan full circles joined with solid and dashed lines represent numerically simulated data for the forward and backward directions, respectively. All other parameters are fixed ($a=3,~b=4$ and $K_2=0.1$).}
    \label{SF_adaptK1}
\end{figure}

In this section, we construct a network of size $N=5000$ with a power-law degree distribution with a minimum degree of $86$ and maximum degree of $145$. We follow the same procedure of global order parameter adaptation. We observe similar behaviour as in the previous case. First, we adapt $r_1$ to a fraction of nodes with only the triadic coupling $K_2$. Figure (\ref{SF_adapt_K2}) shows the transition scenarios with varying pairwise coupling $K_1$. We see that, for partial adaptation, stable states are generated between incoherent and coherent states, which are broadened with an increasing number of adaptive nodes. Finally, for full adaptation, both hystereses vanish. Thus, the paths from explosive to continuous transitions are induced in the system due to order parameter adaptation.
Similarly, we adapted $r_1$ partially to the pairwise coupling $K_1$ in the presence of adaptation with $K_2$. The formation of an explosive path from the tiered path is clearly depicted in Fig. (\ref{SF_adaptK1}). Further, Figs.(\ref{SF_adapt_K2}) and (\ref{SF_adaptK1}) show that the numerical data points fit well with the analytical ones.     
Thus, we have seen that for degree--degree correlated networks, partial order-parameter adaptation with $K_2$ evolves the transition from explosive to continuous, mediated by different kinds of transition paths with double jumps in the forward and backward directions. To verify the generality of these transitions in other types of networks with different topologies, we choose a synthetic random network and proceeded in the same manner.

\section{Random Network}
\begin{figure}[h!]
    \centering
    \includegraphics[width=9cm]{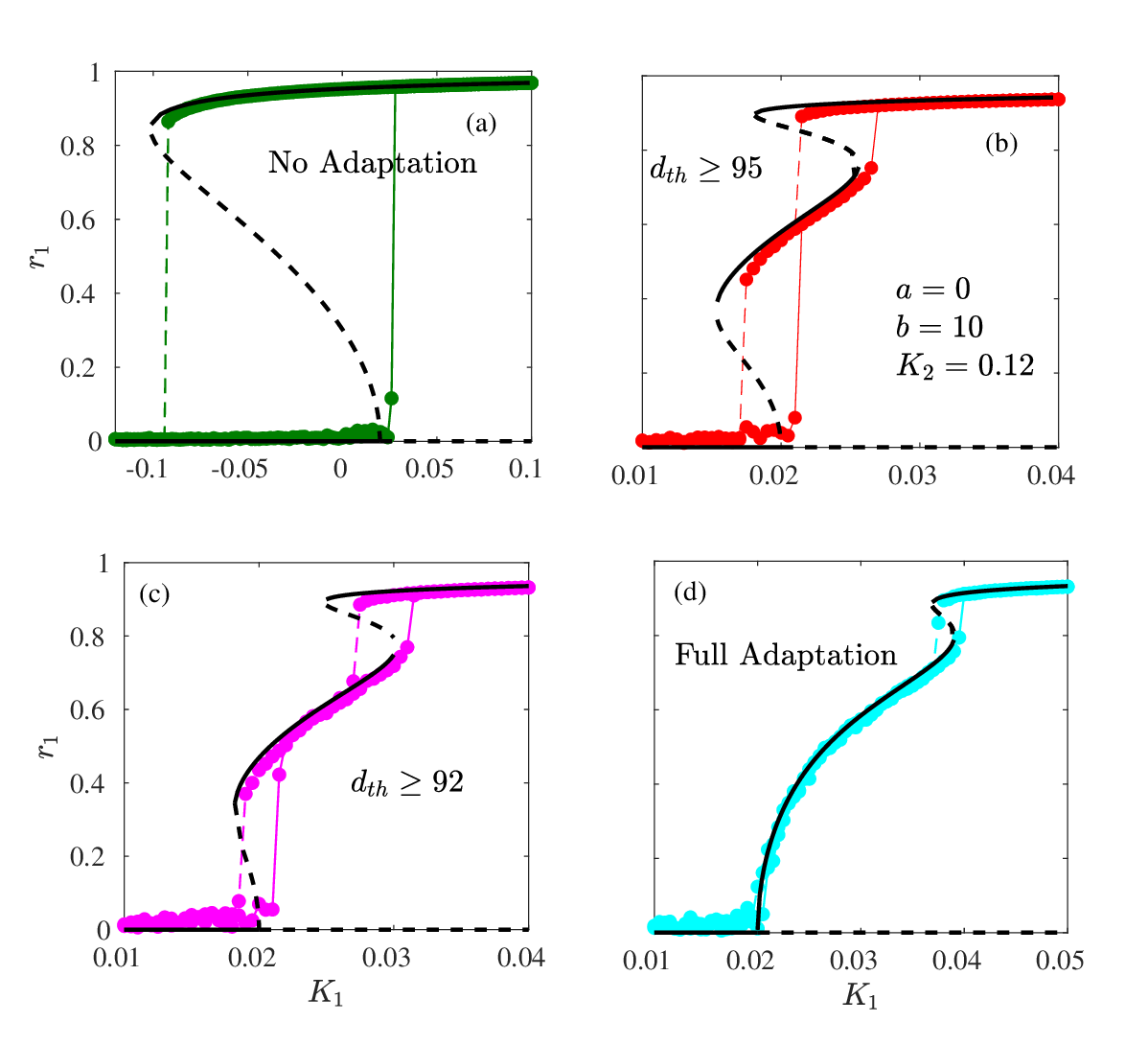}
    \caption{Synchronisation profiles for adapting $r_1$ partially to $K_2$ in a random network. $r_1$ as a function of $K_1$ for (a) no adaptation, (b) $d_{\rm th}=95$, (c) $d_{\rm th}=92$ and (d) full adaptation. Black solid and dashed lines indicate stable and unstable solutions of the self-consistentent equations, respectively. green, red, magenta, and cyan full circles joined with solid and dashed lines represent numerically simulated data for the forward and backward directions, respectively. All other parameters are fixed ($a=0,~b=10$ and $K_2=0.12$).}
    \label{ER_adapt_K2}
\end{figure}

\begin{figure}[h!]
    \centering
    \includegraphics[width=9cm]{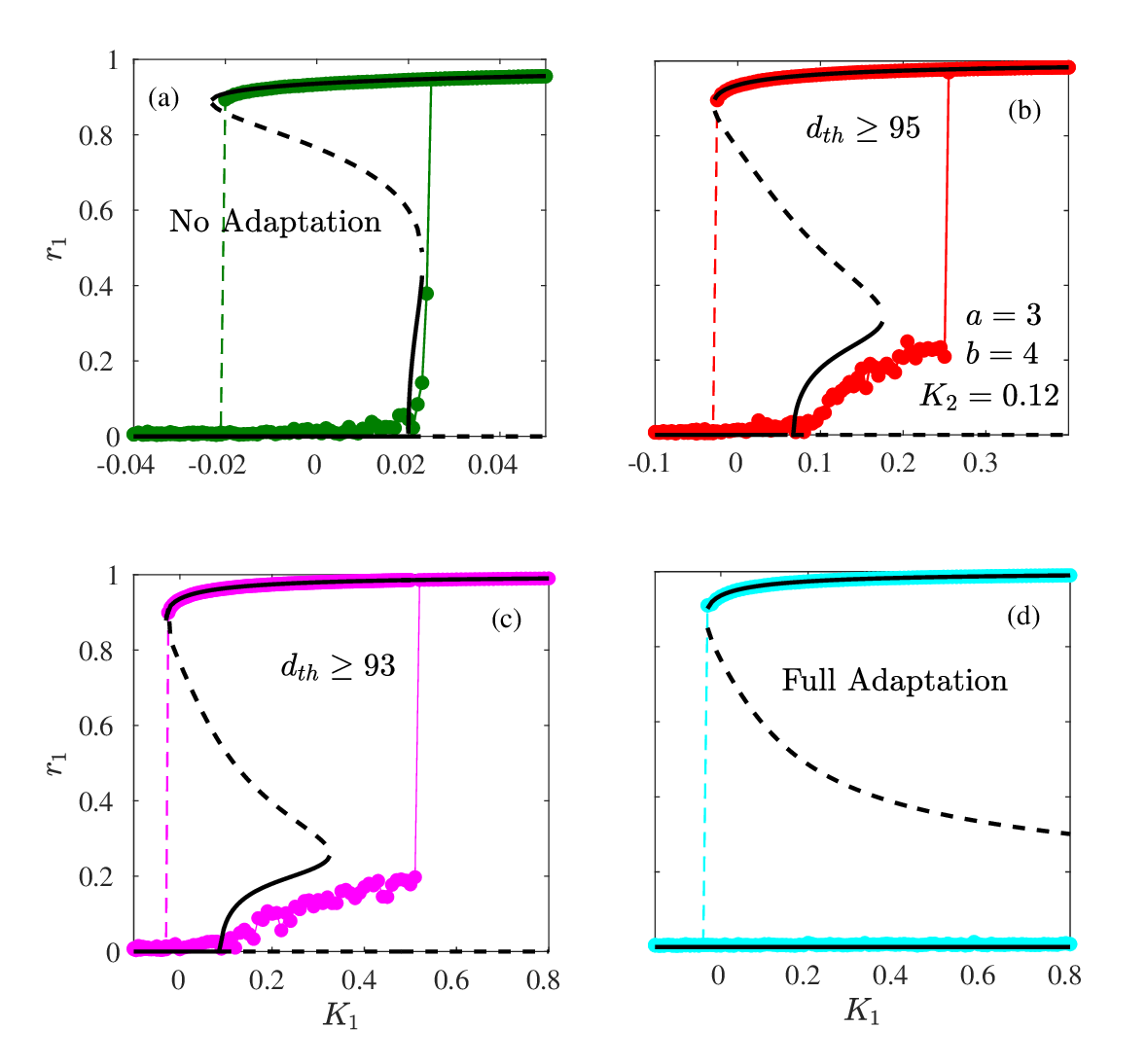}
    \caption{Synchronisation profiles for partially adapting $r_1$ to $K_1$ in a random network. $r_1$ as a function of $K_1$ for (a) no adaptation, (b) $d_{th}=95$, (c) $d_{th}=93$ and (d) full adaptation. Black solid and dashed lines indicate stable and unstable solutions of the self-consistent equations, respectively. Green, red, magenta, and cyan full circles joined with solid and dashed lines represent numerically simulated data for the forward and backward directions, respectively. All other parameters are fixed ($a=3,~b=4$ and $K_2=0.12$).}
    \label{ER_adapt_K1}
\end{figure}

In this section, we check the transition scenarios, with the same technique as above, using a random network. We generated a random network with links created with probability $p^{(2)}(k,k')=\frac{\langle k\rangle}{N}$ (similar to the Erd{\H{o}}s-R{\'e}nyi (ER) network) and triangles created by connecting the nodes with probability $p^{(3)}(k,k',k'')=\frac{2\langle q\rangle}{N^2}$. Here $\langle k\rangle$ and $\langle q\rangle$ denote the mean pairwise degree and mean triangular degree, respectively.
%We consider the same network as described in Sec.(\ref{sec_random_network}) in the main text and 
The steps proceed as we have done in the case of a degree-correlated network. First, inserting the probabilities of creating links and triangles into Eq.\ (\ref{reduced_model_supp}), we obtain
\begin{eqnarray}
    \dot{\alpha}+i\omega_0\alpha+\Delta\alpha-\frac{K_1r_1^a}{2N}\sum_{k'}N(k')\langle k \rangle[\alpha(k')e^{i\beta}&&-\bar{\alpha}(k')\alpha^2(k)e^{-i\beta}]-\frac{K_2r_1^b}{2N^2}\sum_{k',k''}N(k')N(k'') 2\langle q \rangle    \nonumber\\
    &&[\alpha^2(k')\bar{\alpha}(k'')e^{i\beta}-\bar{\alpha}^2(k')\alpha(k'')\alpha^2(k)e^{-i\beta}]=0.   % \label{eq_alpha_er}
\end{eqnarray}
Assuming 
\begin{eqnarray}
    U_1=\frac{1}{N}\sum_{k'}N(k')\alpha(k'), \label{U1_er_supp}\\
    U_2=\frac{1}{N}\sum_{k'}N(k')\alpha^2(k')
    \label{U2_er_supp}
\end{eqnarray}
and proceeding as before, we obtain two similar equations:
% the above equation simplified to 
% \begin{eqnarray}
%     &&\dot{\alpha}+i\omega_0\alpha(k)+\Delta\alpha(k)-\frac{K_1r_1^a\langle k\rangle }{2}[U_1e^{i\beta} -\bar{U}_1\alpha^2(k)e^{-i\beta}] \nonumber\\
%     &&-K_2r_1^b\langle q\rangle[U_2\bar{U}_1e^{i\beta}-\bar{U}_2U_1\alpha^2(k)e^{-i\beta}]=0
% \end{eqnarray}
% Here we seek for stationary rotating solutions of $\alpha$. For that we put the polar forms as
% $\alpha(k,t)=\alpha e^{i\omega_1 t}$, $U_1(t)=U_1e^{i\omega_1 t}$, $U_2(t)=U_2e^{2i\omega_1 t}$. Also we take the modulus of $U_1(t)$ and $U_2(t)$ as $\alpha$ and $\alpha^2$. Then Separating the real and imaginary parts we get 
\begin{eqnarray}
    \Delta \alpha-(\frac{K_1}{2}r_1^a\langle k \rangle U_1\cos\beta+K_2r_1^b\langle q\rangle U_1U_2 \cos\beta) (1-\alpha^2)=0,
    \label{eq_alpha_er_supp}
\end{eqnarray}
\begin{eqnarray}
    \alpha\omega_1 =-\alpha\omega_0+(\frac{K_1}{2}r_1^a\langle k \rangle U_1+K_2r_1^b\langle q\rangle U_1U_2)(1+\alpha^2)\sin\beta.
\end{eqnarray}
Thus, the non-zero solutions of $\alpha$ in Eq.\ (\ref{eq_alpha_er_supp}) combined with Eqs. (\ref{U1_er_supp}) and (\ref{U2_er_supp}) yield the self-consistent equations: 
\begin{eqnarray}
    U_1=\frac{1}{N}\sum_{k}N(k)\alpha(k,U_1,U_2), \\
    U_2=\frac{1}{N}\sum_{k}N(k)\alpha^2(k,U_1,U_2),
\end{eqnarray}
where 
%\begin{scriptsize}
\begin{eqnarray}
    \alpha=\frac{-1+\sqrt{1+(K_1r_1^a\langle k\rangle U_1+2K_2r_1^b\langle q\rangle U_1U_2)^2\cos^2\beta}}{(K_1r_1^a\langle k\rangle U_1+2K_2r_1^b\langle q\rangle U_1U_2)\cos\beta},
\end{eqnarray}
%\end{scriptsize}
with $r_1=U_1$ and $r_2=U_2U_1$.
The solution of these self-consistent equations provides a complete set of order parameter ($r_1$) values. 
We also follow the same procedure of partially adapting the global order parameter to both $K_2$ and $K_1$ 
in a system of size $N=5000$ 
and both the mean degrees equal to 
$100$. 
This process allows us to control the respective coupling strengths. In the case of adapting $r_1$ with $K_2$, we fix the other parameters to $a=0,~b=10$ and $K_2=0.12$ and plot the synchronisation profiles both analytically and numerically. Analogous to Fig. 1 (in the main text), Fig. (\ref{ER_adapt_K2}) demonstrates synchronisation transitions for different threshold values. Clearly, in the absence of $r_1$ adaptation, the system follows an explosive path. Then in Fig. \ref{ER_adapt_K2}(b) the effective $K_2$ is partially reduced for $d_{\rm th}=95$. Because of this, $K_1$ attempts to induce continuous transition paths in the system. Hence a stable state is generated between two unstable states. This type of path allows the system to jump twice. Once again, a decrease in the $d_{th}$ value reduces the widths of both hystereses. Note that in this random network, a small hysteresis remains in the upper part with full adaptation because of the higher $K_2$ value. For the case of non-zero adaptation with 
$K_1$, the synchronisation routes are transformed from tiered to explosive paths (Fig.\ \ref{ER_adapt_K1}). This follows the same trend as in Fig.\ \ref{UD_adaptK1}, i.e. the length of the lower continuous path becomes larger with a smaller $r_1$ value and eventually overlaps with the $r_1=0$ branch. Thereafter, the route becomes explosive.

These investigations prove the scalability of the proposed technique to networks of different configurations. Further, we can conclude from the above observations that the explosiveness of the transition to synchronisation can be controlled through the triadic coupling strength in a networked system. Moreover, with the variation in dominance of $K_1$ and $K_2$ for different parameter values, the networks show different types of transitions, such as explosive, double jump, tiered and continuous.
\section{Evolution of Frequency}
\begin{figure}[h!]
    \centering
    \includegraphics[width=9cm]{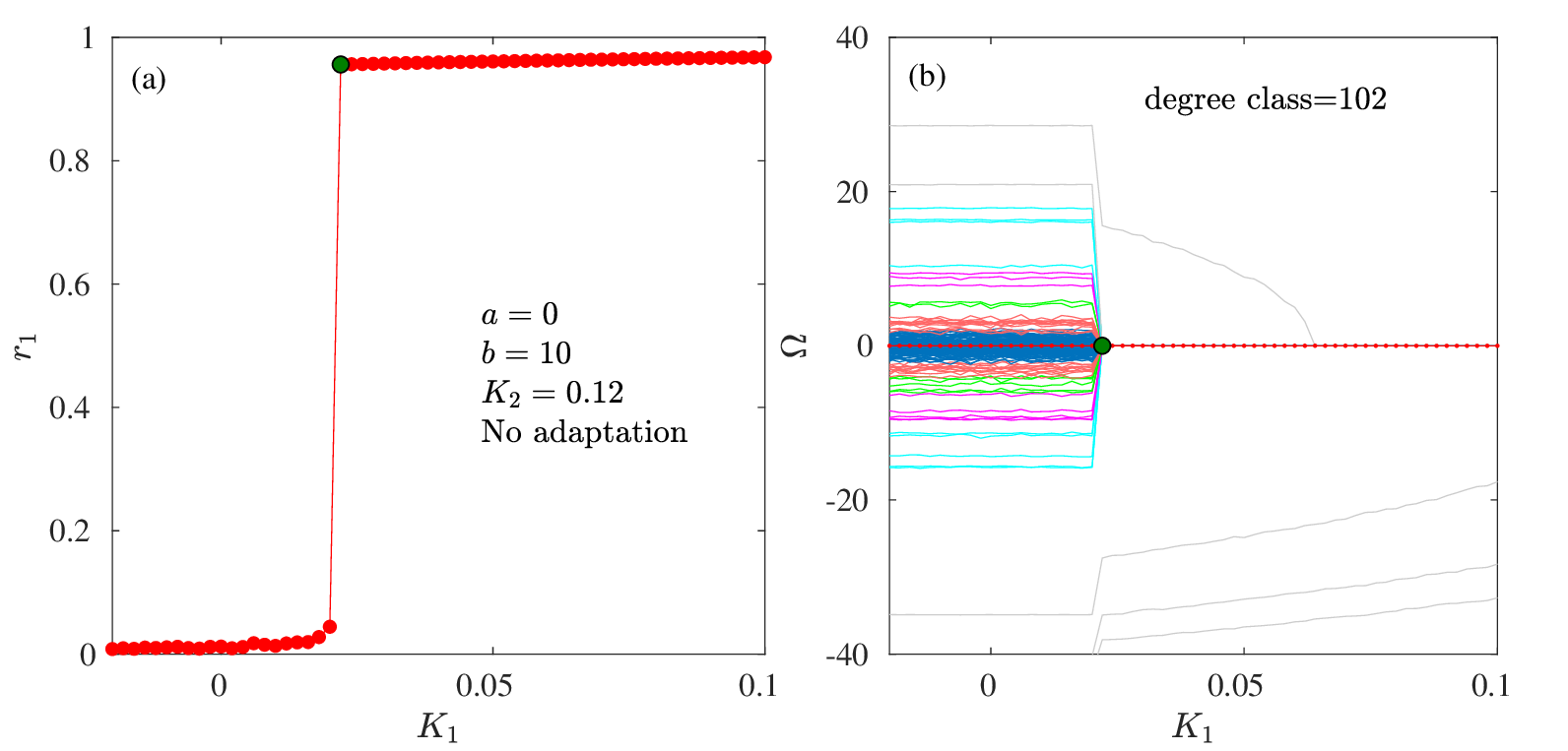}
    \caption{Frequency evolution for corresponding synchronisation transition. (a) $r_1$ as a function of $K_1$ 
    for a single jump in the forward direction. The transition point is marked with a full green circle. (b) Mean frequency as a function of $K_1$ for degree class $102$. Blue, light-red, light-green, magenta, cyan and gray colours denote the frequencies of nodes with natural frequencies in the ranges $|\omega_i|<2,~ (2,4),~(4,6),~ (6,10),~ (10,20)$ and $>20$, respectively. 
    The dotted red line represents the mean frequency of the network.}
    \label{ER_frequency_explosive}
\end{figure}

\begin{figure}[h!]
    \centering
    \includegraphics[width=9cm]{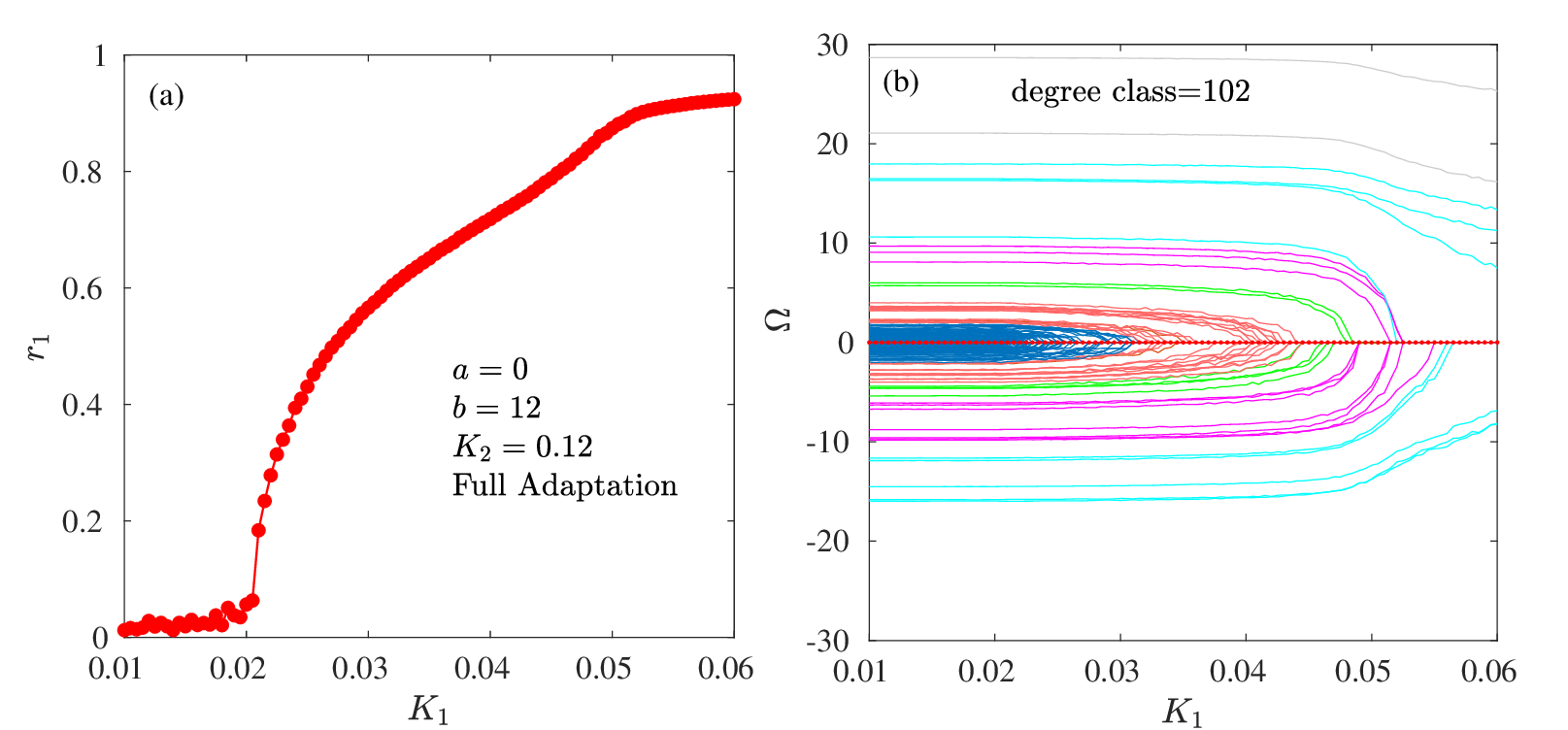}
    \caption{Frequency evolution for corresponding synchronization transition. (a) $r_1$ as a function of $K_1$ for a continuous transition. (b) Mean frequency as a function of $K_1$ for degree class $102$. Blue, light-red, light-green, magenta, cyan and gray colours denote the frequencies of nodes with natural frequencies in the ranges $|\omega_i|<2,~ (2,4),~(4,6),~ (6,10),~ (10,20)$ and $>20$, respectively. The dotted red line represents the mean frequency of the network.}
    \label{ER_frequency_tiered}
\end{figure}

The frequency evolution of the nodes concerning double jumps has been analysed in the main text. To examine the frequency evolution of other synchronisation transitions, we plotted the frequencies as a function of the coupling strength $K_1$. We computed the mean frequency values $\langle \dot{\theta} \rangle$ in forward continuation for an ER graph. Figure (\ref{ER_frequency_explosive}) shows the synchronisation transition and corresponding frequency transition for a single jump in the forward direction. In this case, the nodes with an approximate natural frequency of $|\omega_i|\leq 18$ abruptly jump to the mean frequency at the critical coupling. This signifies that the nodes in the range $|\omega_i|\leq 18$ suddenly join a single cluster. With an increase in the coupling strength, the remaining oscillators join clusters according to their natural frequency, implying an increase in the synchronisation level.    
Similarly, we computed the frequency evolution of the continuous synchronisation transition in the forward direction. Figure \ref{ER_frequency_tiered}(a) shows that the $r_1$ value increases continuously until reaching the synchronised state. Figure \ref{ER_frequency_tiered}(b) shows the frequency transition at the node level. Clearly, as the coupling strength is increased, the mean frequency of the network attracts the nodes according to their natural frequencies. 

% Similarly we have computed the frequency evolution for tiered synchronization transition in forward direction. As seen in Fig.\ref{ER_frequency_tiered}(a), that the $r_1$ value first increases continuously, then at a certain point it abruptly jump to the synchronized state. Figure \ref{ER_frequency_tiered}(b) displays the frequency transition at node level. Clearly the nodes having frequencies in the range $|\omega_i|\leq 4$ continuously converge to the mean frequency $\Omega$ and the nodes in $|\omega_i|\leq 6$ abruptly jump to $\Omega$ at the transition point.  As the coupling strength is increased the mean frequency $\Omega$ attracts the remaining nodes.
\begin{figure}[h!]
    \centering
    \includegraphics[width=9cm]{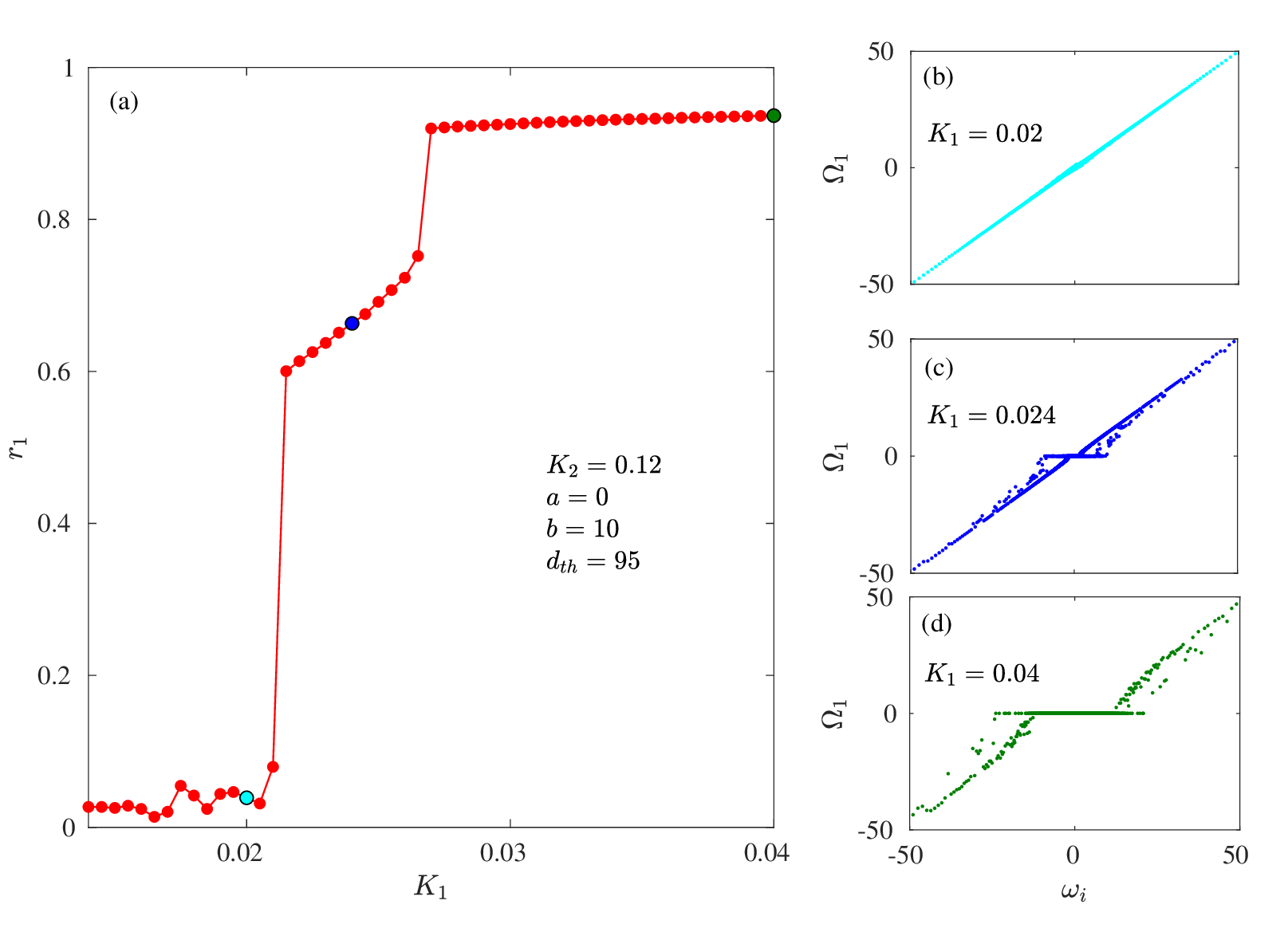}
    \caption{Frequency evolution for corresponding synchronization transition. (a) $r_1$ as a function of $K_1$ for a double-jump transition in forward direction. Cyan, blue and green circles denote three points in different synchronisation states. Mean frequency as a function of natural frequency $\omega$ at (b) $K_1=0.02$ (incoherent state), (c) $K_1=0.024$ (partially synchronized state), (d) $K_1=0.04$ (synchronized state). }
    \label{ER_full_frequency}
\end{figure}

From the above discussion, we have elucidated the microscopic dynamics of the considered system and the basic mechanisms behind each type of synchronisation transition. Now, we can aim to understand the macroscopic dynamics of the whole system. In Fig. \ref{ER_full_frequency} we have plotted the mean frequency ($\Omega_1$) as a function of natural frequency ($\omega_i$) for different coupling strengths. Figure \ref{ER_full_frequency}(b) shows $\Omega_1$ at $K_1=0.02$. Here, the mean frequencies are spread along a line, implying that the system is in an incoherent state (Fig. \ref{ER_full_frequency}(a)). At $K_1=0.024$, nodes with $|\omega_i|< 10 $ gather at the $\Omega_1=0$ line (Fig. \ref{ER_full_frequency}(c)). This characterises the emergence of a single cluster. Further, the noise around the $\Omega_1=0$ line indicates that the nodes with these frequency values will merge at that cluster if the $K_1$ value is increased. Finally at the synchronised state ($K_1=0.04$), the nodes with higher frequencies merge with the $\Omega_1=0$ line (Fig.\ref{ER_full_frequency}(d)), implying that the size of that single cluster has increased. Moreover, this size will increase with the coupling strength.    
\newpage
%\bibliographystyle{apsrev4-1}
%%\bibliography{Pitu_master_bib}
%\bibliography{References1}

%

\end{document}